\documentclass[aps,prx,twocolumn,superscriptaddress,nofootinbib]{revtex4-2}

\usepackage{amsmath}
\usepackage{graphicx}
\usepackage{amssymb}
\usepackage{color}
\usepackage{mathrsfs}
\usepackage{verbatim} 
\usepackage[colorlinks,linkcolor=blue,urlcolor=black,citecolor=blue,anchorcolor=blue]{hyperref}
\usepackage[T1]{fontenc}
\usepackage{orcidlink}
\usepackage{float}

\begin{document}	
\title{Genuine and spurious (non-)ergodicity in single particle tracking}

\author{Wei Wang~\orcidlink{0000-0002-1786-3932}}
\thanks{These authors contributed equally to this work}
\affiliation{University of Potsdam, Institute of Physics \& Astronomy, 14476
Potsdam-Golm, Germany}
\author{Qing Wei~\orcidlink{0000-0001-5287-9430}}
\thanks{These authors contributed equally to this work}
\affiliation{LSEC, ICMSEC, Academy of Mathematics and Systems Science, Chinese
Academy of Sciences, Beijing 100190, China}
\affiliation{Max-Planck-Institut f{\"u}r Physik komplexer Systeme, N{\"o}thnitzer Straße 38, 01187 Dresden, Germany}
\author{Igor M. Sokolov~\orcidlink{0000-0002-4688-9162}}
\email{igor.sokolov@physik.hu-berlin.de}
\affiliation{Institut f{\"u}r Physik, Humboldt-Universit{\"a}t zu Berlin, Newtonstraße 15, D-12489 Berlin, Germany}
\author{Ralf Metzler~\orcidlink{0000-0002-6013-7020}}
\email{rmetzler@uni-potsdam.de}
\affiliation{University of Potsdam, Institute of Physics \& Astronomy, 14476
Potsdam-Golm, Germany}
\affiliation{Asia Pacific Centre for Theoretical Physics, Pohang 37673,
Republic of Korea}
\author{Aleksei Chechkin~\orcidlink{0000-0002-3803-1174}}
\email{achechkin@mpi-halle.mpg.de}
%\affiliation{University of Potsdam, Institute of Physics \& Astronomy, 14476
%Potsdam-Golm, Germany}
\affiliation{Asia Pacific Centre for Theoretical Physics, Pohang 37673,
Republic of Korea}
\affiliation{Max Planck Institute of
Microstructure Physics, Weinberg 2, 06120 Halle (Saale), Germany}
\affiliation{Faculty of Pure and Applied Mathematics, Hugo Steinhaus Center,
Wroc{\l}aw University of Science and Technology, Wyspianskiego 27, 50-370
Wroc{\l}aw, Poland}
\affiliation{Akhiezer Institute for Theoretical Physics, National Science Center,
Kharkiv Institute of Physics and Technology, Akademichna 1, Kharkiv 61108,
Ukraine}

\begin{abstract}
In single-particle tracking experiments measuring anomalous diffusion dynamics,
understanding ergodicity is crucial, as it ensures that the time average
of an observable matches the ensemble average---and can thus be fitted with
known ensemble-averaged observables. A commonly used criterion for assessing
the ergodicity of a stochastic process is based on the comparison of the mean-squared
displacement (MSD) with the time-averaged MSD (TAMSD).  This approach has
been widely applied and proves effective in cases of weak ergodicity breaking
across various systems in both theoretical and experimental studies. However,
there is relatively little discussion regarding the theoretical justification
and limitations of this definition. Here, we demonstrate that this widely
accepted criterion to some extent contradicts the classical definition of
ergodicity as well as physical intuition, leading to spurious (non-)ergodicity
results when applied to several well-known stochastic models. To address this
limitation, we propose using the mean-squared increment (MSI) instead of
the MSD for comparison of ensemble- and time-averaged observables. For processes with stationary increments, the results for MSI and MSD coincide. However, if the increments of the process reach stationarity at long times only, they differ, and the MSI-TAMSD comparison still provides the test for ergodicity. Several
well-established examples demonstrate, through both trajectory simulations and theoretical analysis, that our MSI-TAMSD criterion 
%not only
%effectively reveals weak ergodicity breaking, equivalent to the MSD-TAMSD
%approach, but also 
provides a more accurate characterization of the genuine
(non-)ergodicity of systems where the MSD-TAMSD method fails. Additionally,
for systems exhibiting "ultraweak" ergodicity breaking, the MSI can reveal the
asymptotic stationarity and ergodic nature of the process' increments. Our
findings emphasize the important role of the MSI observable for SPT experiments
and anomalous diffusion studies.

%The MSI is, in fact, equivalent
%to Kolmogorov's structure function---a measure that is capable of unveiling aging and
%stationary properties, as originally introduced  to describe
%locally homogeneous and isotropic turbulence.
\end{abstract}
%\tableofcontents
\maketitle
	
\section{Introduction}

Single particle tracking (SPT) enables the precise recording of the motion
of micron-sized or even smaller fluorescently-labeled tracer particles such
as biomolecules, vesicles, virions, molecular complexes, or colloids. SPT has been
established as a powerful tool to probe dynamic processes and transport in
complex environments such as biomembranes or the cytoplasm of living biological
cells \cite{elena2020}, thus providing access to statistical characterization
of such processes \cite{manzo2015,barkai2012}. SPT has been extensively
applied to investigate protein, DNA, and
granule motion within cellular compartments \cite{bhatia2016,liang2014,ized2014,
jeon2011,mend2024,christine,tabei}, vesicle and viral particle transport
\cite{ewers2005,yosh2006,zhuang2003,cherstvy2019}, and protein dynamics
within membranes \cite{jeon2016,manzo2014,simo1995,simons2010,elena2013}.

One of the most central results obtained from SPT in such complex systems is
the observation and measurement of the anomalous diffusion, i.e., diffusive motion
deviating from ordinary Brownian motion (BM). Anomalous diffusion is quite
ubiquitous and typically characterized by a power-law form of the
mean-squared displacement (MSD) \cite{barkai2012,metz2014a,franosch2013} 
%\footnote{In this paper, for simplicity and without loss of generality, we consider random processes without drift, i.e., $\left<x(t)\right>=0$.}
\begin{eqnarray}\label{eq-msd}
\mathrm{MSD}(t)=\left<[x(t)-x(0)]^2\right>\sim 2K_{\alpha}t^{\alpha},
\end{eqnarray}
where $K_{\alpha}$ is the generalized diffusion coefficient of physical
dimension $[K_{\alpha}]=\mathrm{length}^2/\mathrm{time}^{\alpha}$, 
$\alpha$ is the anomalous diffusion exponent, $x(0)$ is the initial position of the particles, and $\left<\cdot\right>$ represents the ensemble average. In this paper, for simplicity and without loss of generality, we consider random processes without drift, i.e., $\left<x(t)\right>=0$. When $0<\alpha<1$, the process
is called subdiffusive, often arising from trapping, crowding, or obstructed
environments. When $\alpha>1$, the motion is referred to as superdiffusion,
typically driven by positive, long-range correlations (long memory) or large
jumps (L{\'e}vy walks). Ordinary Brownian motion corresponds to normal (Fickian)
diffusion with $\alpha=1$. Biologically relevant examples of subdiffusion at
the micron-scale are usually associated with impeded motion of tracers in the
crowded cytoplasm of living biological cells \cite{weiss2004,gold2006,jeon2011,
mend2024,christine,tabei}, in artificially crowded environments \cite{banks2005,
szym2009,weis2021,lene1}, in entangled actin networks \cite{wong2004,yael1,
krapf2017}, in concentrated protein solutions \cite{pan2009}, as well as in
lipid bilayer systems \cite{jeon20121,akim2011,eiji,gupta,he}. Superdiffusion
has been observed, inter alia, for many active motor-driven transport processes in
living cells \cite{caspi2000,song2018,chen2015,christine}, cell motion and cargo
transport on carpets of active cells \cite{beta,beta1}, for trajectories of
serotonergic fibers in the brain \cite{janusonis2020,janusonis2023}, and for the
motion of animals \cite{jelt23prr,vilk20221}. Outside of biology, enhanced diffusion is observed in cosmic rays \cite{ucha2013,yan2014}, in soft Lorentz gases \cite{klages2019}
and in quantum many-body systems \cite{wei2022}.

Different stochastic mechanisms may give rise to the same MSD scaling while the other statistical characteristics are fundamentally distinct. For example,  anomalous diffusion processes with similar MSD behavior may have different probability density functions (PDFs) and correlation functions, or may or may not show aging \cite{metz2014a,soko2012,meroz2011}. %statistical features, 
%such as aging, stationarity, or ergodicity. 
Therefore, a reliable interpretation of SPT data requires not only identifying the diffusion exponent, but also assessing other properties of the process. 
%trajectory reproducibility and whether time-averaged observables are representative of their ensemble-averaged counterparts. 
Model classification of an anomalous diffusion process is a central goal of the recent community-wide AnDi project \cite{andi,andi2}. 

One of the standard approaches for model classification uses a comparison of the MSD, Eq.~(\ref{eq-msd}), with its time-averaged counterpart called time-averaged
mean-squared displacement (TAMSD),
%analogue In SPT experiments involving diffusion phenomena studies, the main observable is the displacement $x(t+\Delta)-x(t)$, i.e., the increment of the process $x(t)$. The notion of ergodicity is generalized to such two-time variables, and most interest is paid on the behavior of the squared increment. The statistics corresponding to the left hand side of Eq.~(\ref{eq-ergodicity}) is now the time-averaged
%mean-squared displacement (TAMSD) 
\begin{equation}\label{eq-tamsd}
\mathrm{TAMSD}(\Delta,T)=\frac{1}{T-\Delta}\int_0^{T-\Delta}\left[x(t'+
\Delta)-x(t')\right]^2dt',
\end{equation}
considered as a function of the "lag time" $\Delta$  and averaged along a single trajectory of length $T$.
In this context, $\Delta$ is considered as an analogue of $t$ in Eq.~(\ref{eq-msd}). In SPT studies, comparing the $\mathrm{MSD}(t)$ with $\mathrm{TAMSD}(t,T)$ in a long-$T$ limit has been widely used to identify and distinguish different anomalous diffusion processes in both theoretical models and experimental systems \cite{metz2014a,soko2015}.
The equivalence of both was considered as a sign of "ergodicity",
%(in a somewhat overstretched sense, as we proceed to show),
while the discrepancies between the two are often interpreted as signatures of ergodicity breaking \cite{he2008,soko2008}.  
Here we show that while this concept is an important concept in that it creates awareness that time and ensemble averaged quantities may differ from each other and thus care needs to be taken for parameter regression from data, a more careful approach is needed when talking about (non)-ergodicity of a given system.

\begin{figure}
\hspace*{-0.8cm}
\includegraphics[width=1.2\linewidth]{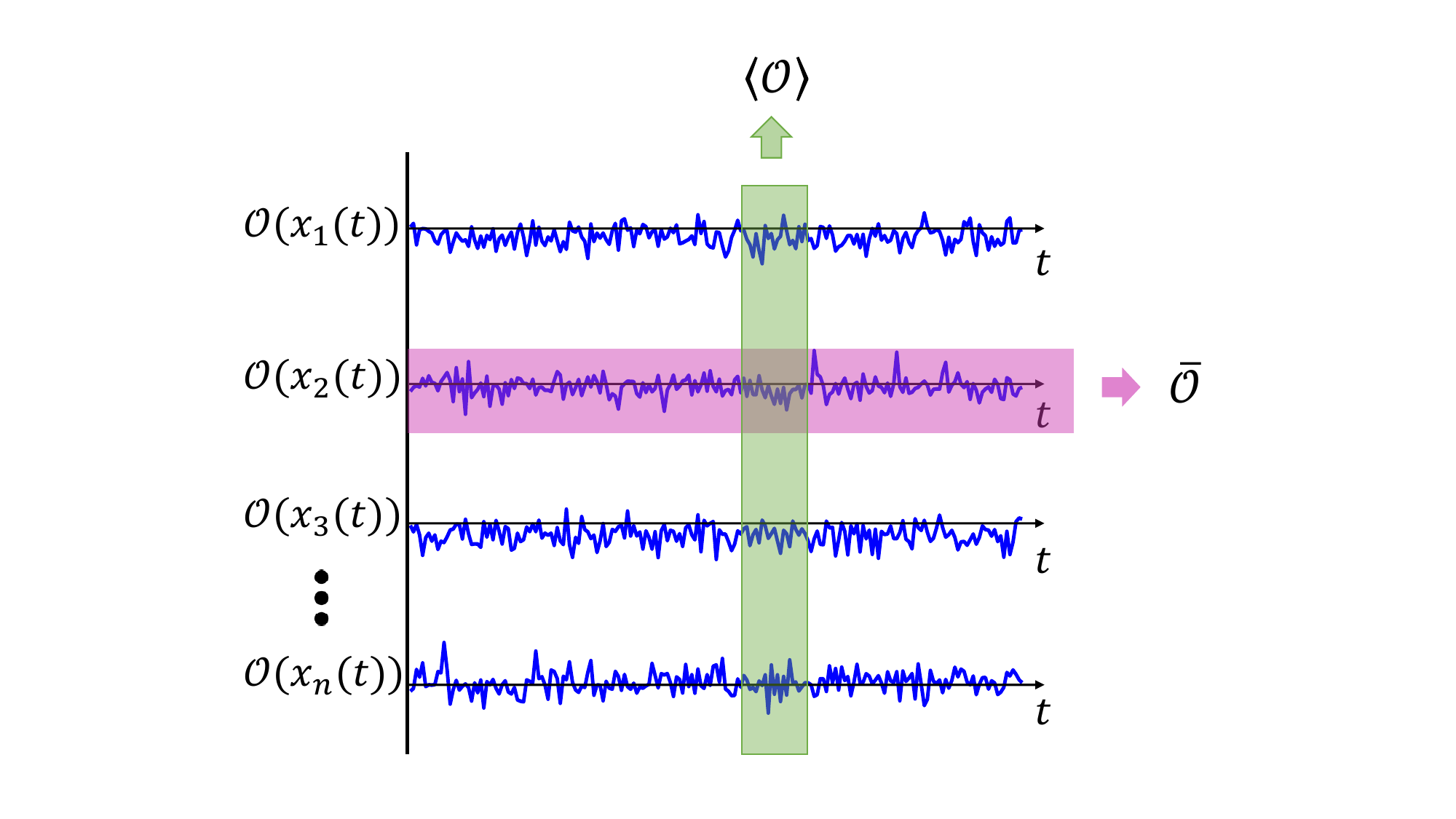}
\caption{Schematic for the ergodicity of an observable $\mathcal{O}$ of a
stochastic process $x(t)$ with stationary PDF $P_{\mathrm{st}}(x)$. The
ensemble average $\left<\mathcal{O}\right>=\int_{-\infty}^\infty\mathcal{O}
(x)P_{\mathrm{st}}(x)dx$ and the time average $\overline{\mathcal{O}}=\frac{1
}{T}\int_0^T{\mathcal{O}(x(t'))}dt'$ of one single trajectory are highlighted
by magenta and green colors respectively. The process $x(t)$ is ergodic with
respect to $\mathcal{O}$ if $\lim_{T\to\infty}\overline{\mathcal{O}}=\left<
\mathcal{O}\right>$.}
\label{fig-schematic}
\end{figure}

Ergodicity is a fundamental concept in the theory of dynamical systems and
stochastic processes. The term "ergodic" originates from the Greek words "ergon"
(work) and "odos" (path), and was originally introduced by Boltzmann in his
ergodic hypothesis within the kinetic theory of gases in statistical mechanics
\cite{walters1982}. The formalization of this concept came through the ergodic
theorem,  developed by Birkhoff \cite{birk1931}, Khinchin \cite{khinchin} and von Neumann
\cite{neumann1932}, initiating a new field of mathematical research called
ergodic theory---that has thrived ever since and was of great significance in
various fields, including mathematics \cite{ergotheory,khinchin}, physics
\cite{arnold}, economics \cite{peters2019}, and social sciences
\cite{fisher2018}. Essentially, ergodicity is an asymptotic property which asserts that for an asymptotically stationary
stochastic process $x(t)$, the time average of an observable quantity
$\mathcal{O}(x)$ over an infinite time span converges to its ensemble average,
\begin{eqnarray}
\label{eq-ergodicity}
\underbrace{\lim_{T\to\infty}\frac{1}{T}\int_0^T\mathcal{O}(x(t'))\, dt'}_{
\mbox{Time average }\overline{\mathcal{O}}}=\underbrace{\int_{-\infty}^\infty
\mathcal{O}(x)P_{\mathrm{st}}(x)\, dx}_{\mbox{Ensemble average }\langle
\mathcal{O}\rangle},
\end{eqnarray}
where $P_{\mathrm{st}}(x)$ is the \textit{stationary} PDF assigning weights to $x$. If the limit on the left-hand-side of Eq.~(\ref{eq-ergodicity}) exists, it must be time-independent, and so must be the right-hand-side.  Thus all ergodic
processes are stationary processes (but not necessarily vice versa). 
For instance,
Gaussian noise and L{\'e}vy noise qualify as both stationary and ergodic
\cite{magd2018}. In contrast, certain special stochastic processes, e.g.,
$X(t)=C$, where $C$ is a random variable, are stationary but not ergodic
processes.
%There are two main types of stationarity
%\cite{ross1995}: strict stationarity (strong stationarity), i.e., the $n$-point 
%PDFs of the processes are invariant under time shifts, and wide-sense stationarity (second-order
%or weak stationarity), i.e., the processes have a time-invariant mean, and the
%covariance between observations depends only on the time difference. In SPT
%experiments, the full $n$-point PDF is typically unknown; therefore, weak
%stationarity is commonly considered. 
If Eq.~(\ref{eq-ergodicity}) holds,
averaging over repeated experiments in an ensemble of similar systems yields the same results as a time average
over a single, but prolonged experimental run, as shown by the schematic
in Fig.~\ref{fig-schematic}.

Here we distinguish between ergodicity in dynamical systems and in stochastic processes. In dynamical systems theory, ergodicity refers to whether the system as a whole is ergodic. If it is, any observable (i.e., measure-preserving transformation) will satisfy Eq.~(\ref{eq-ergodicity}). To assess this property, however, one must know both the state space and the underlying dynamics of the system.
In contrast, in the case of stochastic processes in experimental settings, little or nothing is typically known about the full state space of the system, and random processes provide only a reduced description of its dynamics. In this context, Eq.~(\ref{eq-ergodicity}) serves as an operational definition of ergodicity for a specific observable. Since only measurable quantities can be analyzed, ergodicity must be defined separately for each statistical property of interest, for instance, whether a process is ergodic with respect to its mean, second moment, or correlation function.
We also note that Eq.~(\ref{eq-ergodicity}) may not hold when the long-time limit on the left hand side is not defined. For
instance, the corresponding integral may still fluctuate even in the limit
of infinite data acquisition times. This case corresponds to the situation
when the initial condition is never forgotten. Different scenarios may
take place.  One option is that the state space of the system may consist
of several disconnected parts, and starting in each of them leads to a
different value of the time average, a situation called "strong ergodicity
breaking" \cite{bouch1992}. Alternatively, the state space may be connected but the dynamics
on it "freezes", and a relevant trajectory cannot sample it fully even
at infinite time. This situation corresponds to what is called "weak
ergodicity breaking" \cite{bouch1992,metz2014a,burov2011}. Note that these
situations cannot be resolved on the basis of single trajectories only,
and there is not much difference from the point of view of the stochastic
processes whether the system itself is non-ergodic or ergodic but exhibiting
an extremely slow relaxation \cite{soko2015}. Weak ergodicity breaking
is observed in many experimental situations in which systems displaying
aging behavior, traditionally included glassy systems such as spin glasses
\cite{bouch1992},
gels \cite{cipe2000}, nanocrystals \cite{bouch2003,barkai2004}, or quantum
many-body systems \cite{turner,khemani}. Scenarios of strong ergodicity breaking in similar systems are discussed in \cite{bernaschi,roy}.

In SPT experiments involving diffusion phenomena studies, the main observable is the displacement $x(t+\Delta)-x(t)$, i.e., the increment of the process $x(t)$. 
Note that Eq.~(\ref{eq-ergodicity}) typically defines the equality of the ensemble and time averages for an observable depending on the value of $x$ at one single time, while the squared displacements considered in Eqs.~(\ref{eq-msd}) and (\ref{eq-tamsd}) involve the values of $x$ at two times. This fact is indeed of relevance, as we proceed to show. 
The notion of ergodicity is generalized to such two-time variables (by considering a set of increments for a fixed time lag), and most interest is paid to the behavior of the squared increment. 
The statistic corresponding to the left hand side of Eq.~(\ref{eq-ergodicity}) is now the TAMSD, Eq.~(\ref{eq-tamsd}).
 The statistic corresponding to the right hand side of Eq.~(\ref{eq-ergodicity}) can be defined in different ways. The standard way is using the MSD defined by Eq.~(\ref{eq-msd}), which explicitly depends on the initial condition. This approach was applied across various processes, both in theoretical studies
and experiments \cite{wang2022a,jeon2011,barkai2005,diego2011,manzo2015,
diego2017,tabei,vilk2022}, giving the possibility to distinguish between different processes. However, it is not suitable as the test for ergodicity. 
%A standard method for assessing the (non-)ergodicity of a stochastic process is to compare the MSD and TAMSD in the limit of long trajectories, as proposed in \cite{he2008,metz2014a}. 
%The discordance between MSD and TAMSD is then considered as
%a sign of ergodicity breaking. 
%This approach was applied across various processes, both in theoretical studies
%and experiments \cite{wang2022a,jeon2011,barkai2005,diego2011,manzo2015,
%diego2017,tabei,vilk2022}. 
%However, it contradicts the probabilistic definition of  ergodicity (\ref{eq-ergodicity}).
For instance,  for BM, 
MSD-TAMSD equivalence holds,
in spite of the fact that the process is nonstationary and should be classified as nonergodic.
Moreover, the Ornstein-Uhlenbeck process starting at the minimum of the potential, is asymptotically stationary and ergodic, however it yields a TAMSD that is twice the MSD at
long times, as we will see below. 
%According to the MSD-TAMSD comparison, this
%would indicate nonergodicity---an outcome inconsistent with the established
%understanding of the process.

%This is not a suitable observable to be used in "real ergodicity" tests, but still a very useful one.

In this paper, we show through several generic examples which are well-known random processes and commonly
used in the description of SPT experiments, that the widely used criterion based on the comparison of the MSD and TAMSD is not suitable for determining the (non-)ergodic nature of a process.  \textit{In fact, the MSD represents the statistical property of the initial increment $x(t)-x(0)$, whereas, the definition of the TAMSD employs arbitrary increments $x(t+\Delta)-x(t)$ whose statistical properties may be different from those of the initial increment.} 
%In fact, the equivalence of the MSD and TAMSD implies stationarity of the increments of the process. 
Using the mean-squared increment
(MSI) instead of the MSD, that is
\begin{equation}\label{eq-msi}
\mathrm{MSI}(t,\Delta)=\left<[x(t+\Delta)-x(t)]^2\right>,
\end{equation}
offers a robust and consistent framework for
evaluating genuine (non-)ergodicity. Especially, when the process has stationary increments, the MSI reduces to
the MSD. However, if the process reaches stationarity at long times only, the MSD is different from the MSI, with the latter being equal to the TAMSD if the increment process is ergodic.

The MSI is essentially the structure function introduced by Kolmogorov in his study of Gaussian, self-similar processes with stationary, power-law-correlated increments—commonly referred to as Wiener spirals \cite{kolmogorov1940}. Such processes have been employed in the analysis of locally homogeneous and isotropic turbulence \cite{kolmogorov1941a,kolmogorov1941b}, providing a statistical characterization of turbulent flows.
%Building on similar statistical principles,
%Kolmogorov also described a class of Gaussian, self-similar processes
%with stationary, power-law-correlated increments---known as Wiener spirals
%\cite{kolmogorov1940}. 
Kolmogorov's pioneering work laid the foundation for later
advancements, particularly the 1968 work by Mandelbrot and Van Ness in which
they formulated an explicit integral representation of these processes---termed
fractional Brownian motion (FBM) \cite{mandelbrot1968}. Today, FBM is
recognized as a canonical example of a Gaussian process with long-range
correlations, characterized by the Hurst exponent $H$ \cite{mishura2008};
for $H=1/2$, FBM reduces to classical BM. Notably, FBM exhibits stationary
and ergodic increments (FBM itself is nonstationary and thus nonergodic)---properties that can be
effectively evaluated through the MSI and TAMSD. \textit{In contrast, similar to BM, the equivalence of
MSD and TAMSD traditionally used in the SPT framework, falsely suggests ergodicity for FBM} \cite{metz2014a}. This example highlights the
limitations of the traditional viewpoint and demonstrates the advantages of an MSI-based analysis for
investigating ergodic behavior.

In particular, the MSI is instrumental in understanding the phenomenon of
"ultraweak ergodicity breaking" \cite{godec2013,jeon2023,wei2025}, where a
process demonstrates the same scaling behavior in both the MSD and TAMSD,
yet with differing prefactors (see also \cite{zumofen}). While this mismatch
breaks the MSD-TAMSD equivalence---signaling that the process itself is
nonergodic---the increments can still exhibit asymptotic stationarity at
long times, as revealed by the MSI for FBM in the form of a Riemann-Liouville
fractional integral (RL-FBM) \cite{lim2001}, and for L{\'e}vy walks
\cite{froemberg2013a}.

The outline of this paper is as follows. Sec.~\ref{section2} introduces the
fundamental physical observables of stochastic processes frequently employed
in SPT experiments and presents both the traditional MSD-TAMSD criterion
and the proposed MSI-TAMSD criterion for evaluating (non-)ergodicity.
Secs.~\ref{section3} to \ref{section6} examine several well-established
stochastic models frequently encountered in theoretical and experimental
studies, demonstrating the limitations of the traditional MSD-TAMSD approach,
which may lead to spurious (non-)ergodicity, and highlighting the robustness
and effectiveness of our proposed MSI-TAMSD criterion. Specifically,
Sec.~\ref{section3} shows that the conventional MSD-TAMSD equivalence falsely
indicates ergodicity in BM and FBM, while MSI-TAMSD accurately captures
the ergodic nature of their increments. Sec.~\ref{section4} explores
cases of spurious nonergodicity in the Ornstein-Uhlenbeck process (OUP) and
diffusion processes under resetting---both are in fact stationary and ergodic
processes. Sec.~\ref{section5} addresses the ultraweak ergodicity breaking in
RL-FBM and L{\'e}vy walks---here the MSI-TAMSD criterion can reveal
the asymptotic stationarity and ergodicity of the increments in these
systems. In Sec.~\ref{section6}, the MSI-TAMSD criterion is applied to
the continuous-time random walk (CTRW) model to assess the genuine
ergodicity and nonergodicity in the cases with exponential waiting time PDF
and power-law waiting time PDF, respectively. Finally, in Sec.~\ref{section7} and \ref{section8},
we summarize and discuss our results.

\section{Main observables of stochastic processes}
\label{section2}

SPT provides the entire trajectory of the tracked particle which allows one to
evaluate different observables to quantify the dynamics of the system under
observation. The most central measurable to quantify the averaged diffusive
behavior of tracer particles is the MSD, defined through the mathematical expectation as
\begin{equation}
\left<\Omega^2(t)\right>=\int_{-\infty}^\infty \int_{-\infty}^\infty(x-x_0)^2
P(x,t|x_0,0)P_0(x_0)dxdx_0.
\end{equation}
Here $P(x,t|x_0,0)$ is the propagator (Green's function) of the process and
$P_0(x_0)$ is the PDF of the initial position $x_0$.

The corresponding empirical MSD is obtained by averaging the time series
$x_i(t)$ over ensembles of trajectories (labeled with the subscript $i$) at a
given time $t$, with respect to each trajectory's initial position,
\begin{eqnarray}\label{eq-empirical-msd}
\nonumber
\left\langle\Omega^2(t)\right\rangle_{\mathrm{emp}}&=&\left<[x(t)-x(0)]^2\right>_{\mathrm{emp}}\\
&=&\frac{1}{N}\sum_{i=1}^N[x_i(t)-x_i(0)]^2.
\end{eqnarray}
Here $N$ is the total number of trajectories. The number $N$ of trajectories should be chosen as large as possible, but sometimes is limited by experimental conditions.

The MSI, qualifying the increment of displacement
\begin{eqnarray}
\delta(\Delta;t)=x(t+\Delta)-x(t)
\end{eqnarray}
during the lag time $\Delta$ starting at physical time $t$, is defined as the
mean-squared of the increment in the form \cite{gikhman2004}
\begin{equation}\label{eq-msi}
\left<\delta^2(\Delta;t)\right>=\left<[x(t+\Delta)-x(t)]^2\right>.
\end{equation}
In analogy to the empirical MSD, the empirical MSI is obtained by averaging over $N$ trajectories,
\begin{eqnarray}
\left<\delta^2(\Delta;t)\right>_{\mathrm{emp}}
=\frac{1}{N}\sum_{i=1}^N [x_i(t+\Delta)-x_i(t)]^2.
\end{eqnarray}
The MSI is equal to the MSD $\left<\Omega^2(\Delta)\right>$ if the process has
stationary increments. The definition of the MSI is identical to the structure
function originally introduced by Kolmogorov and later employed by Yaglom in their works on
locally homogeneous and isotropic turbulence \cite{yaglom1953,yaglom1955,
yaglom1987,kolmogorov1941a,kolmogorov1941b}.

The diffusive dynamics of an individual particle can be also quantified from a
single particle trajectory $x(t)$ via the TAMSD \cite{barkai2012,metz2014a,
he2008}
\begin{equation}
\label{tamsd}
\overline{\delta^2(\Delta;T)}=\frac{1}{T-\Delta}\int_0^{T-\Delta}\left[x(t'+
\Delta)-x(t')\right]^2dt',
\end{equation}
where $T$ is the length of the time series (measurement time) and $\Delta$ is
the lag time. The TAMSD is typically used to evaluate few but long time series
garnered in SPT experiments, e.g., in biological cells, of geo-tagged larger
animals, or of financial time series \cite{manzo2015,barkai2012,nathan2022,
chers2017}. We note that we use the term "TAMSD" for historical reasons. \textit{In fact, the TAMSD also concentrates on increments and more
precisely, 
this quantity represents the time-averaged squared-increment.}

Based on the TAMSD one can obtain the ensemble-averaged TAMSD (EA-TAMSD or mean
TAMSD) by averaging over an ensemble of $N$ individual trajectories in the form
\cite{he2008,barkai2012,metz2014a}
\begin{eqnarray}\label{eq-ea-tamsd}
\left<\overline{\delta^2(\Delta;T)}\right>_{\mathrm{emp}}=\frac{1}{N}\sum_{i=1}^N
\overline{\delta_{ i}^2(\Delta;T)}.
\end{eqnarray}

A routine comparison criterion for a given time series in experiments is the equivalence of the MSD and the TAMSD in the limit of
long trajectories $\Delta/T\ll 1$,
\begin{equation}
\label{eq-msd-tamsd}
\overline{\delta^2(\Delta;T)}\sim \left<\Omega^2(\Delta)
\right>_{\mathrm{emp}}.
\end{equation} 
This relation is frequently employed to assess the ergodicity of stochastic
processes \cite{metz2014a,he2008,barkai2012}, and its violation is commonly
interpreted as an indication of weak ergodicity breaking. While this MSD-TAMSD
criterion (\ref{eq-msd-tamsd}) has been successfully applied in a large number
of systems revealing weak ergodicity breaking phenomena, it contradicts the
classical definition of ergodicity given by (\ref{eq-ergodicity}) and physical
intuition. Especially, when applied to nonstationary processes with stationary
increments, such as BM and FBM, the equivalence (\ref{eq-msd-tamsd}) of MSD and
TAMSD may lead to spurious ergodicity. Conversely, it may wrongly indicate
nonergodicity in systems that are actually stationary and ergodic, such as the
OUP, as shown below. Moreover, unlike the classical definition (\ref{eq-ergodicity}), the
MSD-TAMSD criterion (\ref{eq-msd-tamsd}) compares different physical observables
on each side of the equation, further weakening its theoretical justification.
The difference between the ensemble average (\ref{eq-empirical-msd}) and the moving window average (\ref{tamsd}), is in fact due to the nonstationarity of the process
under consideration.

We note that in the definition (\ref{tamsd}) of the TAMSD, the moving time
window method essentially tests the properties of the increments of the
process at different times, and ergodicity breaking can have its source either
in the nonstationarity of the increments, or in the fact that different trajectories of the process, whether with stationary increments or not, are
intrinsically very different from one another. Instead, in line with the
classical definition of ergodicity given by relation (\ref{eq-ergodicity}), we
propose to employ a different comparison criterion to test the ergodic nature
of a process: the equivalence between the MSI and the TAMSD in SPT experiments for sufficiently long trajectories ($T$ large) in the long-time limit ($t$ large),
that is,
\begin{equation}
\label{eq-msi-tamsd} 
\overline{\delta^2(\Delta;T)}\sim\left<
\delta^2(\Delta;t)\right>_{\mathrm{emp}}. 
\end{equation}
This relation, central to this work, describes the ergodicity of the increments $\delta(
\Delta;t)=x(t+\Delta)-x(t)$.  In theoretical calculations,
we should begin by checking the equivalence
\begin{equation}
\lim\limits_{\Delta/T\to 0}\left<\overline{\delta^2(\Delta;T)}\right>=\lim_{\Delta/t\to 0}\left<\delta^2(\Delta;t)\right>
\end{equation}
between the ensemble average of the TAMSD (\ref{tamsd}) and the MSI (\ref{eq-msi}). In the
next step we then evaluate the ergodicity breaking (EB) parameter
\cite{he2008,metz2014a,cherstvy2013,barkai2012}
\begin{eqnarray}
\label{eq-eb}
\mathrm{EB}(\Delta;T)=\frac{\left<\left(\overline{\delta^2(\Delta;T)}\right)^2
\right>-\left<\overline{\delta^2(\Delta;T)}\right>^2}{\left<\overline{
\delta^2(\Delta;T)}\right>^2},
\end{eqnarray}
which measures the variance of the TAMSD amplitudes for a given combination
of $\Delta$ and $T$. The EB parameter vanishes as $T\to\infty$ for processes
whose increments are ergodic. As detailed below in the example of  the subdiffusive CTRW
with a scale-free waiting time PDF, EB converges to a finite value in the
limit $T\to\infty$. For further details on the EB parameter in both BM and
FBM, we refer the reader to Appendix~\ref{app-a}. We also note that in some
literature instead of EB parameter, the closely related relative standard deviation of
the TAMSD or the magnitude correlation function are employed
\cite{takuma,takuma1,takuma2}.

In the following, by examining several well-established stochastic models frequently
encountered in theoretical and experimental studies, we demonstrate that
the MSI-TAMSD criterion (\ref{eq-msi-tamsd}) offers broader applicability for assessing the
(non-)ergodic nature of stochastic processes, effectively eliminating the
spurious (non-)ergodicity indicated by the MSD-TAMSD criterion (\ref{eq-msd-tamsd}).

\section{Spurious ergodicity}
\label{section3}

Let us start by considering FBM, a long-range dependent Gaussian process 
that can be defined in terms of the Langevin equation driven by an external random force exhibiting power-law
correlations (memory),
\begin{equation}
\label{MNFBM}
\frac{d}{dt}x(t)=\sqrt{2K_{2H}}\xi_{H}(t).
\end{equation}
Here $\xi_{H}(t)$ is fractional Gaussian noise, which is a stationary
Gaussian process with zero-mean and long-time power-law decay of the
ACVF, $\left<\xi_{H}(t)\xi_{H}(t')\right>\sim H(
2H-1)|t-t'|^{2H-2}$ ($|t-t'|\to\infty$). Here we used the Hurst exponent
$H$. The latter is connected to the anomalous diffusion exponent via $\alpha
=2H$, and its range is $H\in(0,1]$. Ordinary BM corresponds to the
limiting case $\alpha=1$ in which the correlations vanish \cite{mandelbrot1968,
qian2003}.

Unconfined FBM itself is nonstationary and thus nonergodic, but its sequence
of increments is stationary and ergodic \cite{deng2009}. The ergodicity of  fractional Gaussian noise is a consequence of Khinchin’s theorem stating that any stationary Gaussian process
is ergodic if its ACVF decays to zero at infinity \cite{
ito1944}. 

For FBM,  the MSD, MSI and EA-TAMSD, are identical, i.e.,
\cite{wei2025}
\begin{eqnarray}
\label{eq-ergodicity-bm}
\left<\Omega^2(\Delta)\right>=\left<\delta^2(\Delta;t)\right>=\left<
\overline{\delta^2(\Delta;T)}\right>=2K_{2H}\Delta^{2H}.
\end{eqnarray}
The equivalence between the MSD and EA-TAMSD incorrectly suggests ergodicity
of the process itself, since FBM is nonstationary and thus nonergodic. Instead, examining the equivalence between the MSI and the EA-TAMSD, together with the vanishing variance of the TAMSD in the limit $T\to\infty$—that is, the EB parameter approaching zero \cite{deng2009}—confirms the validity of Eq.~(\ref{eq-msi-tamsd}). This demonstrates that the increments of these processes exhibit ergodic behavior.
Further details on the EB parameter for FBM are provided in Appendix~\ref{app-a}.

Memory effects can also originate from the viscoelastic properties of the medium. For example, in the cytoplasm of a biological cell, a particle moves through a crowded environment containing elastic structures, where the cytoplasm “pushes back”, generating long-time correlations in the particle’s trajectory \cite{weber}. In this case, the particle exhibits subdiffusive
FBM, which can be described by the fractional Langevin equation (FLE) \cite{lutz2001}.
Analogous to FBM, the FLE has identical MSD, MSI and EA-TAMSD. Therefore it
has ergodic increments \cite{deng2009}, which can be revealed by the equivalence between MSI and EA-TAMSD, even though the process itself governed by the FLE  is nonergodic.

\section{Spurious ergodicity breaking}
\label{section4}

Consider a process $x(t)$ with a specific initial condition $x(0)=0$ that becomes asymptotically
stationary after a characteristic time $\tau$. In the stationary state, the second moment of $x(t)$
attains a constant value denoted by $\langle x^2\rangle_{\mathrm{st}}$. The EA-TAMSD
can then be rewritten in the form
\begin{eqnarray}
\nonumber
\left<\overline{\delta^2(\Delta;T)}\right>&=&\frac{1}{T-\Delta} \int_0^{\tau}
\left<\left[x(t'+\Delta)-x(t')\right]^2\right>dt'\\
\nonumber
&+&\frac{1}{T-\Delta}\int_\tau^{T-\Delta}\left<\left[x(t'+\Delta)-x(t')
\right]^2\right>dt'\\
&\approx&2\left<x^2\right>_{\mathrm{st}},
\label{tamsdou}
\end{eqnarray}
where in the last step we assumed that $\tau\ll\Delta\ll T$. In this limit
the first term in the top line of Eq.~(\ref{tamsdou}) vanishes, and the second
term corresponds to the stationary state. This result indicates that the TAMSD becomes twice the MSD, which equals the second moment under the condition $x(0)=0$ in the stationary state, thereby leading to a spurious nonergodic behavior,
according to the MSD-TAMSD criterion (\ref{eq-msd-tamsd}). In this section,
we study two prototype ergodic systems: the OUP and diffusion with stochastic resetting. Both the OUP and diffusion with stochastic resetting converge to a stationary state. We demonstrate that the MSI can reproduce
the ergodicity of such processes, via the comparison (\ref{eq-msi-tamsd}) with
the EA-TAMSD. The corresponding EB parameter is discussed in Appendix~\ref{app-a}.

\subsection{Ornstein-Uhlenbeck process}

\subsubsection{Normal Ornstein-Uhlenbeck process}

The OUP is one of the most fundamental stochastic processes, with widespread
application in financial mathematics and the physical sciences \cite{kampen1992,
zwanzig2001}. Originally proposed to describe the velocity distribution and
relaxation of a Brownian particle subject to a friction force, the OUP is a Gaussian
and Markovian process. It exhibits a stationary state and is
ergodic. The process is governed by the overdamped Langevin equation \cite{uhlenbeck1930,
gardiner2004}
\begin{eqnarray}
\label{eq-ou}
\frac{dx}{dt}=-\lambda x+\sigma\xi(t),
\end{eqnarray}
with the initial condition $x_0$. $\xi(t)$ is white Gaussian
noise---the time derivative of Brownian motion (or Wiener process)---with zero mean and ACVF $\left<\xi(t)\xi(t')\right>=\delta(t-t')$. The
strength of the restoring force defines the characteristic relaxation time
$\lambda^{-1}$, and $\sigma$ quantifies the noise strength. When $\lambda\to 0$, the OUP reduces to the free BM with diffusion coefficient $K_1=\sigma^2/2$.

Solving the stochastic differential equation (\ref{eq-ou}), $x(t)$ can be
formally expressed as
\begin{equation}
x(t)=e^{-\lambda t}\left(x_0+\sigma\int_0^te^{\lambda s}\xi(s)ds\right).
\end{equation}
When $t\gg\lambda^{-1}$, the stationary state of the OUP follows the
Gaussian stationary distribution
\begin{equation}
\label{eq-ou-distribution}
P_{\mathrm{st}}(x)=\sqrt{\frac{1}{2\pi\left<x^2\right>_{\mathrm{st}}}}\exp
\left(-\frac{x^2}{2\left<x^2\right>_{\mathrm{st}}}\right),
\end{equation}
with the stationary value
\begin{equation}
\label{eq-thermal-value}
\left<x^2\right>_{\mathrm{st}}=\frac{\sigma^2}{2\lambda}
\end{equation}
of the second moment.

The ACVF of the OUP is given by \cite{mardoukhi2020,cherstvy2018}
\begin{equation}
\label{eq-ou-acvf}
\left<x(t_1)x(t_2)\right>=\left(\left<x_0^2\right>-\frac{\sigma^2}{2\lambda}
\right)e^{-\lambda(t_1+t_2)}+\frac{\sigma^2}{2\lambda}e^{-\lambda|t_1-t_2|}.
\end{equation} 
Notably, in the long-time limit $t_1,t_2\gg \lambda^{-1}$, the process becomes asymptotically stationary and thus ergodic. In the case when the initial condition $x_0$ is drawn from the 
stationary
distribution (\ref{eq-ou-distribution}), i.e., $\left<x^2_0\right>=\sigma^2/(2
\lambda)$, the process is ergodic starting from $t=0$ with the ACVF  depending solely on the time difference
$|t_1-t_2|$. 

From expression (\ref{eq-ou-acvf}) for the ACVF, we can derive the MSD,
\begin{equation}
\label{eq-oup-msd}
\left<\Omega^2(t)\right>=\left<x^2_0\right>\left(1-e^{-\lambda t}\right)^2+
\frac{\sigma^2}{2\lambda}\left(1-e^{-2 \lambda t}\right), 
\end{equation}
the MSI 
\begin{eqnarray}
\label{eq-ou-msi}
\nonumber
\left<\delta^2(\Delta;t)\right>&=&\langle\Delta x^2\rangle_{\mathrm{st}}
\left(1-e^{-\lambda \Delta}\right)^2e^{-2\lambda t}\\
&&+\frac{\sigma^2}{\lambda}\left(1-e^{-\lambda \Delta}\right),
\end{eqnarray}
and the EA-TAMSD 
\begin{eqnarray}
\label{eq-ou-tamsd}
\nonumber
\left<\overline{\delta^2(\Delta;T)}\right>&=&\langle\Delta x^2\rangle_{
\mathrm{st}}\left(1-e^{-\lambda\Delta}\right)^2\frac{1-e^{-2\lambda(T-\Delta)
}}{2\lambda(T-\Delta)}\\
&&+\frac{\sigma^2}{\lambda}\left(1-e^{-\lambda \Delta}\right).  
\end{eqnarray} 
In Eqs.~(\ref{eq-ou-msi}) and (\ref{eq-ou-tamsd}), the factor $\langle\Delta
x^2\rangle_{\mathrm{st}}=\left(\left<x^2_0\right>-\sigma^2/[2\lambda]\right)$
measures the difference between the variance of the initial condition and the
stationary state.

In the limit $\lambda\to0$, the OUP reduces to free BM with stationary
increments. In this scenario, the MSD, MSI, and EA-TAMSD are all equivalent,
as given by Eq.~(\ref{eq-ergodicity-bm}) with $H=1/2$ and $K_1=\sigma^2/2$.  

When $\lambda\neq0$, the system behavior depends on the choice of the initial
conditions. If the process is initialized with the  stationary
distribution (\ref{eq-ou-distribution}), i.e., $\left<x^2_0\right>=\sigma^2/
(2\lambda)$, the MSD, MSI, and EA-TAMSD all coincide,
\begin{equation}
\label{eq-OUP-msi-stat-initial}
\left<\Omega^2(\Delta)\right>=\left<\delta^2(\Delta;t)\right>=\left<\overline{\delta^2(
\Delta;T)}\right>=\frac{\sigma^2}{\lambda}\left(1-e^{-\lambda\Delta}\right)
\end{equation}
At long times $\Delta\gg\lambda^{-1}$, all three measures converge exponentially
to the stationary value $\sigma^2/\lambda$, which corresponds to twice the second moment (\ref{eq-thermal-value}) in the stationary state. The equivalence
(\ref{eq-OUP-msi-stat-initial}) indicates that, under stationary initial
conditions, both MSI and MSD reflect the ergodic behavior in
the normal OUP.

\color{black}{This is no longer true when the initial condition is nonstationary}, i.e., $\left<x^2_0\right>\neq \sigma^2/(2\lambda)$. In that
case the system gradually relaxes towards its stationary state. After the time
$t\gg\lambda^{-1}$, the MSD converges to
\begin{equation}
\label{eq-ou-msd2}
\left<\Omega^2(t)\right>\sim\left<x^2_0\right>+\left<x^2\right>_{\mathrm{st}},  
\end{equation}
which explicitly depends on the initial condition. In contrast, in the same
limit $t\gg\lambda^{-1}$ and with $T\gg\lambda^{-1}$, both the MSI
(\ref{eq-ou-msi}) and EA-TAMSD (\ref{eq-ou-tamsd}) exhibit the same
asymptotic behavior
\begin{equation}
\label{eq-ou-msi2}
\left<\delta^2(\Delta;t)\right>\sim\left<\overline{\delta^2(\Delta;T)}\right>
\sim\frac{\sigma^2}{\lambda}\left(1-e^{-\lambda \Delta}\right),
\end{equation}
which thus differ from the MSD (\ref{eq-ou-msd2}). We see that both MSI and
EA-TAMSD are independent of the initial condition and identical to expression
(\ref{eq-OUP-msi-stat-initial}) with stationary initial conditions. In addition,
when $\Delta\gg\lambda^{-1}$, we have
\begin{equation}
\label{eq-ergodicity-ou}
\left<\delta^2(\Delta;t)\right>\sim\left<\overline{\delta^2(\Delta;T)}\right>
\sim 2\left<x^2\right>_{\mathrm{st}}, 
\end{equation}
corresponding to twice the stationary value (\ref{eq-thermal-value}), as already
seen above.

\begin{figure}
(a)\includegraphics[width=0.9\linewidth]{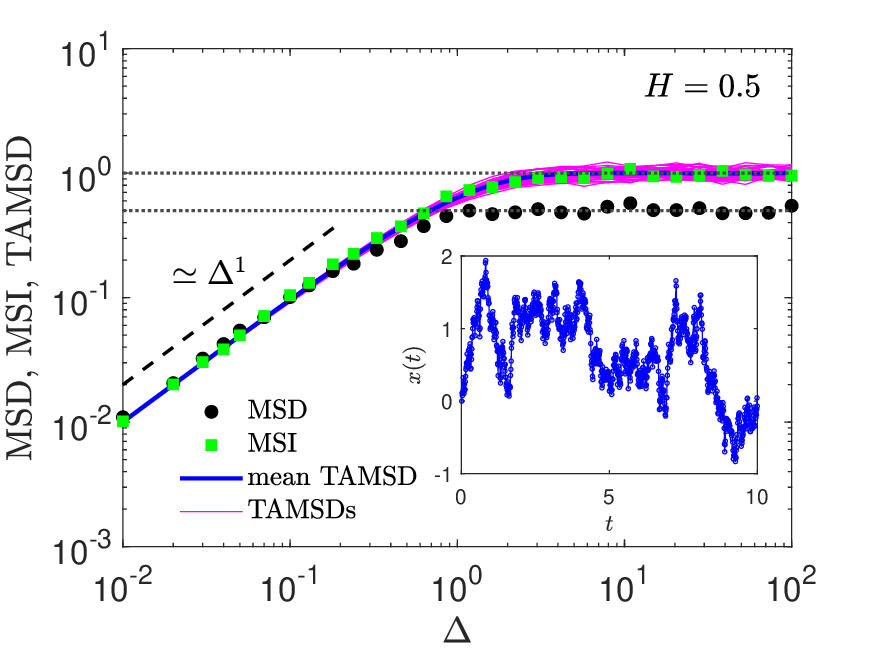}
(b)\includegraphics[width=0.9\linewidth]{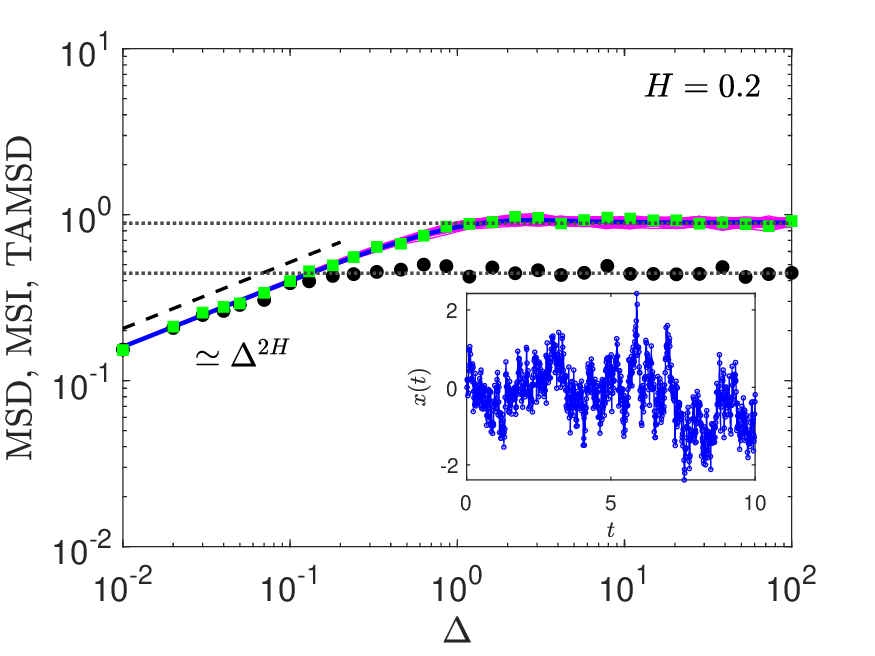}
(c)\includegraphics[width=0.9\linewidth]{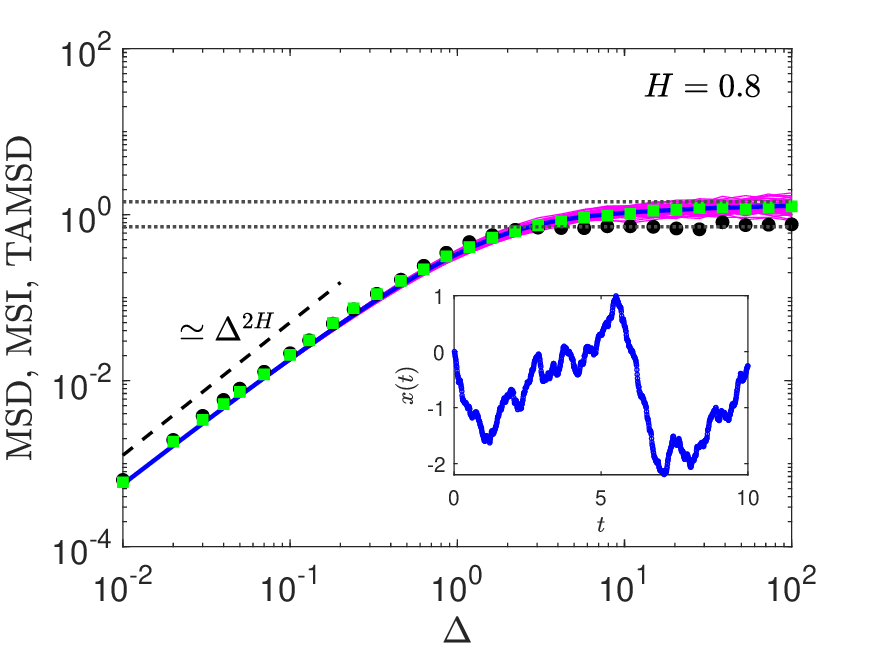}
\caption{Simulation results for Fractional OUP (\ref{eq-foup}) with initial condition $x_0=0$: MSD (black
circles), MSI (green rectangles) and TAMSDs of 50 individual trajectories
(magenta curves) for (a) normal diffusion with $H=0.5$, (b) subdiffusion
with $H=0.2$ and (c) superdiffusion with $H=0.8$. When $H=0.5$, the
fractional OUP reduces to the normal OUP (\ref{eq-ou}) with initial condition $x_0=0$. The EA-TAMSD is
represented by solid blue curves. The variation (scatter) of the amplitudes
among individual TAMSDs is negligible, indicating
that individual trajectories are completely reproducible. The
stationary values of the MSI and MSD are indicated by dotted lines. Single
trajectories are shown in the insets. Parameters: time step $dt=0.01$,
measurement time $T=500$, starting time for the MSI $t=10$, noise strength
$\sigma_{2H}=1$, relaxation rate $\lambda=1$, and number of trajectories
$N=200$.}
\label{fig-foup}
\end{figure}

Interestingly, the stationary value (\ref{eq-ou-msd2}) of the MSD contradicts
the usual intuition that once the process reaches stationarity, all memory of
the initial condition is lost. In contrast, the initial condition variance 
$\left<x^2_0\right>$ does not appear in the limiting value (\ref{eq-ou-msi2})
of TAMSD and MSI. Figure~\ref{fig-foup}(a) presents the MSD, MSI and TAMSDs
of 50 individual trajectories of the normal OUP (\ref{eq-ou}). Although the
OUP is stationary (asymptotically) and ergodic, the fact that the TAMSD (\ref{eq-ou-msi2})
does not match the MSD (\ref{eq-ou-msd2}) when the process arrives at the
stationary state at long times, may lead to a spurious nonergodicity, as
shown in Fig.~\ref{fig-foup}(a). This discrepancy indicates that the MSD is
not a suitable observable for testing ergodicity in this context. However,
as seen from Eqs.~(\ref{eq-ou-msi2}) and (\ref{eq-ergodicity-ou}), the TAMSD
and MSI exhibit the same asymptotic behavior when $t,T\gg\lambda^{-1}$ in both
the short and long lag time regime, making the MSI a more reliable measure for
verifying ergodicity in the OUP. Moreover, the negligible amplitude variation
among the TAMSDs of individual trajectories indicates a high reproducibility,
as shown in Fig.~\ref{fig-foup}(a), reinforcing the ergodic nature of the
process.

\subsubsection{Fractional Ornstein-Uhlenbeck process}

The fractional OUP, i.e., the process (\ref{eq-ou}) driven by fractional
Gaussian noise instead of white Gaussian noise, is an extension of the
normal OUP \cite{oleksii,mardoukhi2020,Cheridito2003}. The fractional OUP is
a Gaussian yet non-Markovian process and has been applied in finance for the
description of a "rough volatility" \cite{gath2018,christ2015}, and in fluid
dynamics to reflect the temporal structure of turbulence \cite{shao1995,
chevill2017}. The fractional OUP can be described by the Langevin equation
\cite{Cheridito2003,mardoukhi2020}
\begin{eqnarray}
\label{eq-foup}
\frac{dx}{dt}=-\lambda x +\sigma_{2H}\xi_H(t),    
\end{eqnarray} where $x_0=x(t=0)$ defines the initial condition, characteristic relaxation time
is $\lambda^{-1}$, $\sigma_{2H}$ quantifies the noise strength, and ${\xi_H}(t)$ denotes
the fractional Gaussian noise introduced above. When $\lambda\to 0$, the fractional OUP reduces to the free FBM with generalized diffusion coefficient $K_{2H}=\sigma_{2H}^2/2$.
In particular, when $H=1/2$,
the fractional OUP reduces to the normal OUP \footnote{Another form of the fractional OUP can be constructed via the Lamperti transformation \cite{lamperti1962}, as detailed in \cite{magd2011}.}.

Similar to the normal OUP, at long times $t\gg\lambda^{-1}$, the fractional OUP
becomes stationary and ergodic \cite{Cheridito2003}, and its PDF follows the
Gaussian distribution
\begin{equation}
\label{eq-foup-distribution}
P_{\mathrm{st}}(x)=\frac{1}{\sqrt{2\pi\left<x_{2H}^2\right>_{\mathrm{st}}}}\exp
\left(-\frac{x^2}{2\left<x_{2H}^2\right>_{\mathrm{st}}}\right),
\end{equation} 
where
\begin{eqnarray}
\left<x_{2H}^2\right>_{\mathrm{st}}=\frac{\sigma_{2H}^2}{2\lambda^{{2H}}}
\Gamma({2H}+1).  
\end{eqnarray}
We note that this stationary value explicitly depends
on the Hurst index, a signature that the fractional OUP is governed by a Langevin equation driven by fractional Gaussian noise, in which the noise is treated as external, as is typical for open systems \cite{klimon1995}. 

The formal solution to the Langevin equation (\ref{eq-foup}) takes the form
\begin{equation}
x(t)=e^{-\lambda t}\left(x_0+\sigma_{2H}\int_0^te^{\lambda s}\xi_H(s)ds\right).
\end{equation}
The ACVF can be expressed in terms of special functions, including the incomplete gamma
function and the Kummer function, as detailed in Appendix~\ref{app-b}. Then, the
MSD is given by
\begin{eqnarray}
\label{eq-foup-msd}
\nonumber
\left<\Omega^2(t)\right>&=&\left<x^2_0\right>\left(1-e^{-\lambda t}\right)^2
+\frac{\sigma_{2H}^2}{2\lambda^{{2H}}}\gamma({2H}+1,\lambda t)\\
\nonumber
&&+\sigma_{2H}^2 t^{{2H}}e^{-\lambda t}-\frac{\lambda \sigma_{2H}^2}{2({2H}+1)}
t^{{2H}+1}e^{-2\lambda t}\\
&&\times\mathcal{M}(2H+1,2H+2,\lambda t),
\end{eqnarray} 
where
\begin{eqnarray}
\gamma(z,x)=\int_0^xe^{-t}t^{z-1}dt
\end{eqnarray}
is the lower incomplete gamma function and 
\begin{equation}
\mathcal{M}(a,b,z)=\frac{\Gamma(b)}{\Gamma(b-a)\Gamma(a)}\int_0^1e^{zt}t^{a-1}
(1-t)^{b-a-1}dt
\end{equation}
is the Kummer function (the confluent hypergeometric function of the first
kind) \cite{abramowitz1964}. In particular, for $H=1/2$, as $M(2,3,\lambda
t)=2[e^{\lambda t}(\lambda t-1)+1]/(\lambda t)^2$ and $\gamma(2,\lambda t)=
1-e^{-\lambda t}(\lambda t+1)$, the MSD reduces to Eq.~(\ref{eq-oup-msd}) of
the normal OUP.  

As we have seen above, the normal OUP is always stationary given the stationary initial conditions. In contrast, the fractional OUP only attains stationarity asymptotically in the long-time limit \cite{mardoukhi2020}, as indicated by its ACVF (\ref{eq-foup-acvf}).

At long times $t\gg\lambda^{-1}$, the MSD (\ref{eq-foup-msd}) of the fractional
OUP approaches a value that depends on the initial conditions,
\begin{eqnarray}
\label{eq-foup-msd2}
\left<\Omega^2(t)\right>\sim\left<x^2_0\right>+\left<x_{2H}^2\right>_{\mathrm{
st}}.
\end{eqnarray}
In contrast, when both $t,T\gg\lambda^{-1}$, the MSI and EA-TAMSD exhibit the
same asymptotic behavior 
\begin{eqnarray}
\label{eq-foup-msi}
\nonumber
\left<\delta^2(\Delta;t)\right>&\sim&\left<\overline{\delta^2(\Delta;T)}\right>\\
\nonumber
&&\hspace*{-2.2cm}
\sim\frac{\sigma_{2H}^2}{\lambda^{{2H}}}\Gamma(2H+1)+\sigma_{2H}^2\Delta^{{2H}}\\
\nonumber
&&\hspace*{-2.2cm}
-\frac{\lambda\sigma_{2H}^2}{2({2H}+1)}\Delta^{{2H}+1}e^{-\lambda\Delta}
\mathcal{M}({2H}+1,{2H}+2,\lambda\Delta)\\
&&\hspace*{-2.2cm}
-\frac{\sigma_{2H}^2}{2\lambda^{{2H}}}\Big\{e^{\lambda\Delta}\Gamma({2H}+1,
\lambda\Delta)+e^{-\lambda\Delta}\Gamma({2H}+1)\Big\}.
\end{eqnarray}
In addition, at long lag times $\Delta\gg\lambda^{-1}$, both the MSI and TAMSD
converge to a value that is independent of the initial condition,
\begin{eqnarray}
\label{eq-foup-msi-tamsd}
\left<\delta^2(\Delta;t)\right>\sim\left<\overline{\delta^2(\Delta;T)}\right>
\sim2\left<x_{2H}^2\right>_{\mathrm{st}}.
\end{eqnarray}
 
Figure~\ref{fig-foup} depicts simulation results for the MSD, MSI, and
TAMSDs of the fractional OUP (\ref{eq-foup}) for different Hurst exponents.
The TAMSDs (\ref{eq-foup-msi-tamsd}) approach stationary values different from the stationary MSD \eqref{eq-foup-msd2}, and from the MSD-TAMSD criterion
\eqref{eq-msd-tamsd} it may lead to the spurious nonergodicity of the
systems. However, the TAMSDs closely follow the behavior of the MSI across
both short and long lag times, indicating the underlying ergodic nature.

A second key difference between normal and fractional OUPs concerns the manner in which
the MSD, MSI, and TAMSD approach their stationary values. In the case of
the normal OUP, all these observables exhibit exponential relaxation, as shown
in Eqs.~(\ref{eq-oup-msd}) and (\ref{eq-ou-msi2}). For the fractional OUP,
however, the dynamics differ significantly, as reported in \cite{jeon2012}
and tested experimentally in \cite{lene1}. Specifically, the MSD with respect to the initial condition $x(0)=0$ relaxes
exponentially (with power-law correction) towards its stationary value, as
shown in Eq.~(\ref{eq-foup-msd-app-st}) of Appendix~\ref{app-b},
\begin{equation}
\left<\Omega^2(\Delta)\right>\sim\left<x_{2H}^2\right>_{\mathrm{st}}-\frac{2
\sigma^2_{2H}}{\lambda^2}H(2H-1)\Delta^{2H-2}e^{-\lambda \Delta}.
\end{equation}
In contrast, the TAMSD and MSI exhibit the same asymptotic behavior (see
Eqs.~\eqref{eq-foup-msi}), and both follow a power-law relaxation toward
stationarity, as shown in Eq.~(\ref{eq-foup-msi-app-st}) in Appendix~\ref{app-b},
\begin{eqnarray}
\nonumber
\left<\delta^2(\Delta;t)\right>&\sim&\left<\overline{\delta^2(\Delta;T)}
\right>\\
&&\hspace*{-1.2cm}
\sim2\left<x_{2H}^2\right>_{\mathrm{st}}-\frac{2\sigma^2_{2H}}{\lambda^2}
H(2H-1)\Delta^{2H-2}.
\end{eqnarray}
This result reinforces the conclusion that the MSI is a more reliable
observable than the MSD for assessing ergodicity in both the normal and
the fractional OUP.

\subsection{Fractional Brownian motion under stochastic resetting}

The simplest form of stochastic resetting involves the diffusion of a single
particle that intermittently returns to its initial position
at a constant, nonzero rate \cite{evans2020}. One of the most prominent effects
of resetting is observed in the position distribution of the particle. Without
resetting, the diffusion process is nonstationary; however, introducing a
nonzero resetting rate leads to a nonequilibrium stationary state (NESS) at
long times \cite{evans2011}. Resetting also significantly influences the
associated first-passage time properties---it has been shown to accelerate
search processes, and an optimally chosen resetting rate can minimize the
mean first-passage time to a target \cite{pal2017,chech2018}. Beyond search
problems, stochastic resetting has a broad range of applications, including 
stochastic thermodynamics \cite{seifert2016}, chemical reaction kinetics
\cite{reuveni2014}, extreme-value statistics \cite{meyl2015}, and the dynamics
of complex networks \cite{boyer2021}. An experimental implementation of
resetting dynamics was demonstrated in \cite{roichman2020}. The (non-)ergodic
properties of systems under stochastic resetting have been explored in recent
studies, for instance, for BM \cite{sand2022,barkai2023}, FBM \cite{wang2022},
and geometric BM \cite{vinod2022,sand2021}. Partial resetting, when the
position is reset to a fraction of its distance to the origin, and incomplete resetting, where the dynamics of the underlying process remain only partly affected by the reset (e.g., when the memory of the process prior to the resetting event is not fully erased), have also been investigated \cite{marcus,shlomi,costantino,costantino1,kacper,bodrova2020}.

Here we consider FBM under Poissonian resetting with the rate $r$ to the origin. Although this model has been shown to reach stationarity, with its ACVF
vanishing at long times \cite{majum2018}, 
we demonstrate that the process is indeed ergodic with respect to the MSI-TAMSD criterion (\ref{eq-msi-tamsd}).
Moreover, our findings can be extended to a broad class of anomalous diffusion
processes under Markovian resetting---where each reset event renews the
system's memory and the waiting times between resets have a finite mean
\cite{wang2022}. We further show that relying solely on the MSD, its
comparison to the TAMSD may lead to misleading conclusions regarding
ergodicity. In contrast, when the MSI is employed instead of the MSD, the ergodic nature of
the process is clearly recovered.

The PDF of FBM under a constant resetting rate $r$ is given by
\cite{maju2015} 
\begin{equation}
\label{eq-reset}
P(x,t)=e^{-rt}P_0(x,t)+\int_0^tre^{-r\tau}P_0(x,\tau)d\tau,
\end{equation} 
where $P_0(x,t)=\exp\left[-x^2/(4K_{2H}t^{{2H}})\right]/\sqrt{4\pi K_{2H}
t^{{2H}}}$ is the free-space propagator describing the probability density
of an FBM particle, initially at the origin $x(0)=0$, to reach position $x$
at time $t$ in the absence of resetting. The first term on the right-hand side
of Eq.~\eqref{eq-reset} corresponds to trajectories, in which no resetting
occurs up to time $t$. The second term accounts for trajectories, that
experience at least one reset before time $t$, with the last reset occurring
at time $t-\tau$, followed by a period of free FBM evolution over the
remaining duration $\tau$. The factor $re^{-rt}$ is the probability density that the last reset occurred at time $t-\tau$, followed by no further resets during the interval of length $\tau$.

Multiplying Eq.~(\ref{eq-reset}) by $x^2$ and performing the integration
with respect to $x$, we obtain the MSD
\begin{equation}
\label{eq-reset-msd}
\left<\Omega^2(t)\right>=2K_{2H}\left[e^{-rt}t^{2H}+\frac{\gamma({2H}+1,rt)}
{r^{2H}}\right],
\end{equation}
where $\gamma(a,t)$ is the lower incomplete gamma function \cite{abramowitz1964}.
When the process becomes stationary at long times $t\gg r^{-1}$, the MSD
saturates to the value
\begin{eqnarray}
\nonumber
\left<\Omega^2(t)\right>&\sim&\frac{2K_{2H}}{r^{2H}}\Gamma({2H}+1)\left(1-\frac{2H}{r}t^{2H-1}e^{-rt}\right)\\
&\sim&\frac{2K_{2H}}{r^{2H}}\Gamma({2H}+1).
\label{eq-reset-msd-lt}
\end{eqnarray}

The exact MSI and EA-TAMSD are presented in the Appendix~\ref{app-c} using the
renewal approach. For $t\gg r^{-1}$ and $T\gg r^{-1}$, the MSI and TAMSD
exhibit the same asymptotic behavior
\begin{eqnarray}
\nonumber
\left<\delta^2(\Delta;t)\right>&\sim&\left\langle\overline{\delta^2(\Delta;T)}
\right\rangle\\
\nonumber
&\sim&2K_{2H}\Bigg[e^{-r\Delta}\Delta^{{2H}}+\frac{\gamma({2H}+1,r\Delta)}{r^{
2H}}\\
&&\hspace*{0.6cm}
+\left(1-e^{-r\Delta}\right)\frac{\Gamma({2H}+1)}{r^{{2H}}}\Bigg].
\label{eq-reset-msi-tamsd1}
\end{eqnarray}
In particular, when $\Delta\gg r^{-1}$ the process approaches the stationary state
exponentially,
\begin{eqnarray}
\nonumber
\left<\delta^2(\Delta;t)\right>&\sim&\left\langle\overline{\delta^2(\Delta;T)}
\right\rangle\\
&&\hspace*{-1.2cm}
\sim4K_{2H}\Bigg[\frac{\Gamma(2H+1)}{r^{2H}}-\frac{H}{r}\Delta^{2H-1}e^{-r\Delta}
\Bigg].
\label{eq-reset-fbm-msi-app}
\end{eqnarray}
Thus, both MSI and EA-TAMSD approach twice the stationary value of the MSD
(\ref{eq-reset-msd-lt}),
\begin{equation}
\label{eq-reset-msi-tamsd2}
\left<\overline{\delta^2(\Delta;T)}\right>\sim\left<\delta^2(\Delta;t)\right>
\sim 2\left<\Omega^2(\Delta)\right>.
\end{equation} 
It is worth noting that, in the long-time limit $\Delta\gg r^{-1}$, the MSD, MSI, and EA-TAMSD all approach their stationary values exponentially with a power-law correction, following the form $\Delta^{2H-1}e^{-r\Delta}$, as
shown in Eqs.~(\ref{eq-reset-msd-lt}) and (\ref{eq-reset-fbm-msi-app}). 

Additionally, we verify the EB parameter through simulations, which vanishes in the long-time limit of the trajectory length $T$ (see Fig.~\ref{eb-reset}).
%The EB has the approximation $\mathrm{EB}(\Delta=\delta t;T)\simeq T^{-1}$ for
%all $H$, which is different from the form of free FBM \cite{deng2009}.
This again emphasizes a key point: the stationary MSD (\ref{eq-reset-msd-lt}) is not a reliable indicator of ergodicity when compared with the TAMSD (\ref{eq-reset-msi-tamsd2}). By contrast, the MSI faithfully captures the ergodic property, as shown in Eq.~(\ref{eq-reset-msi-tamsd2}).

\begin{figure}
(a)\includegraphics[width=1\linewidth]{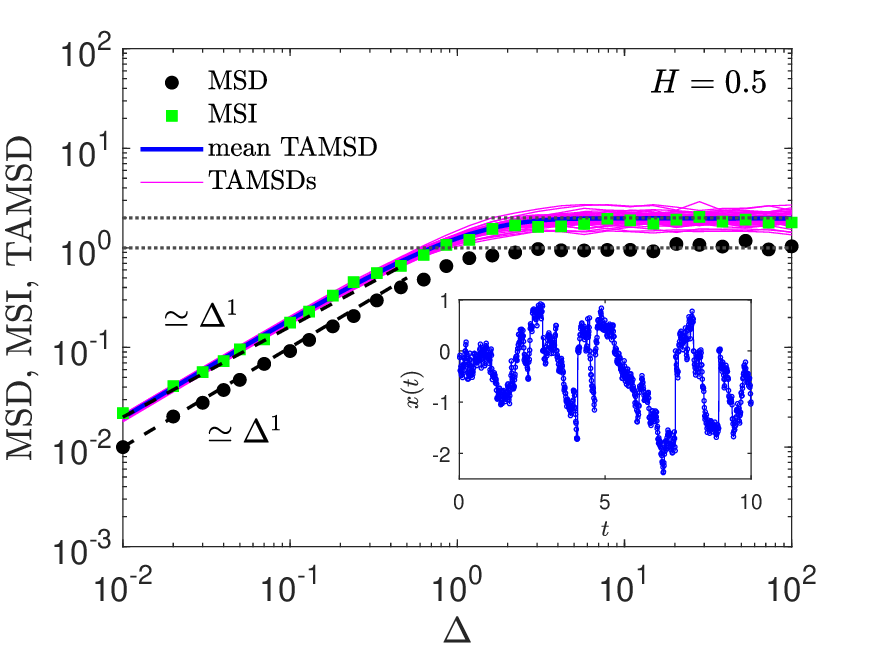}
(b)\includegraphics[width=1\linewidth]{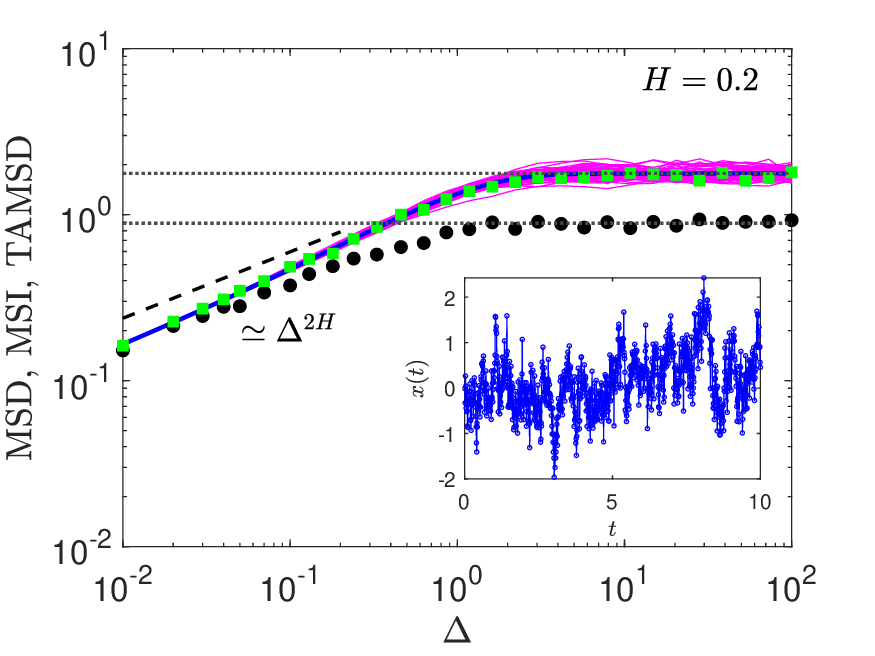}
(c)\includegraphics[width=1\linewidth]{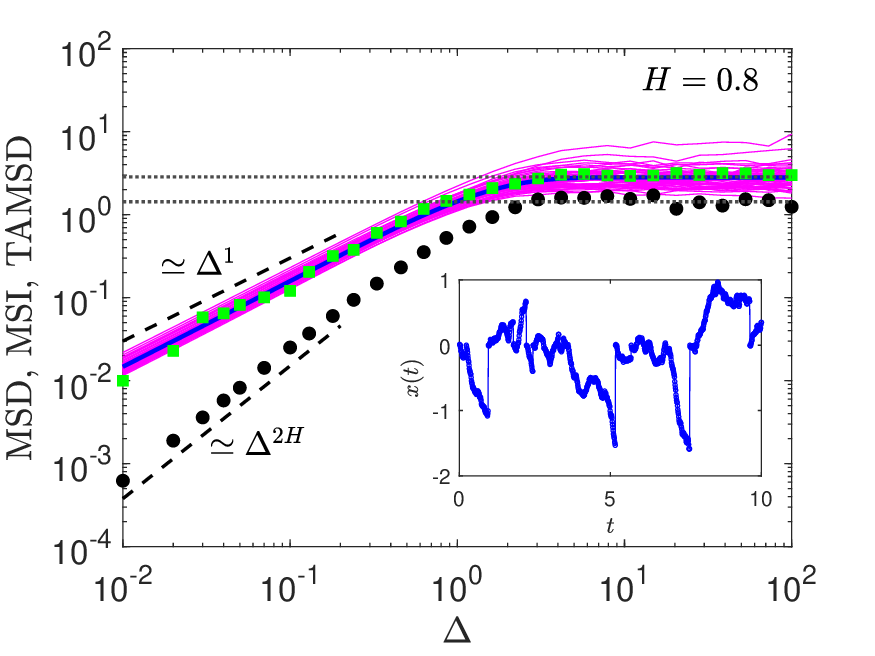}
\caption{Simulation results for FBM under Poissonian resetting with the rate $r$ to the initial position $x(0)=0$, showing MSD (black
circles), MSI (green rectangles) and TAMSDs of individual trajectories (magenta
curves) with (a) $H=0.5$, (b) $H=0.2$ and (c) $H=0.8$. The EA-TAMSD is
represented by solid blue curves. Other parameters: time step $dt=0.01$,
measurement time $T=500$, starting time of the MSI $t=10$, and resetting rate
$r=1$.}
\label{fig-reset}
\end{figure}

Figure~\ref{fig-reset} displays the simulation results for reset FBM and is in
perfect agreement with our theory. We note that for $H\ge1/2$, the resetting
events are clearly visible in the trajectories, whereas for $H< 1/2$, these
events are obscured by the strong fluctuations. 

For both normal and fractional OUPs, at short lag times $\Delta\ll \lambda^{-1}$, the processes are dominated by free BM and FBM, so that the MSD and MSI (or TAMSD) coincide. However, as shown in Fig.~\ref{fig-reset}(a) and (c), for $H\ge1/2$, noticeable deviations between the MSD and MSI (TAMSD) arise at short times $\Delta
\ll r^{-1}$. This discrepancy can be
explained from the short-time approximation (\ref{eq-reset-msi-app2})
for the MSI and TAMSD,
\begin{eqnarray}
\nonumber
\left<\delta^2(\Delta;t)\right>&\sim&\left\langle\overline{\delta^2(\Delta;T)}
\right\rangle\\
&\sim&2K_{2H}\left(\Delta^{2H}+\frac{\Gamma({2H}+1)}{r^{{2H}-1}}\Delta^1\right).
\label{eq-reset-msi-st}
\end{eqnarray}
Unlike the MSD, which remains unaffected by resetting at short times and
follows $\left<\Omega^2(\Delta)\right>\sim2K_{2H}\Delta^{2H}$, for MSI and
TAMSD, however, the linear term in Eq.~\eqref{eq-reset-msi-st} dominates when
$H>1/2$, resulting in a different scaling behavior. When $H=1/2$, both terms
contribute equally, leading the MSI and TAMSD to be approximately twice the
value of the MSD at short times.

\section{Spurious ultraweak ergodicity breaking}
\label{section5}

"Ultraweak ergodicity breaking" refers to the phenomenon in which the MSD
shares the same scaling behavior with the EA-TAMSD, but the two differ by a
constant prefactor, thus violating the equivalence (\ref{eq-msd-tamsd}) of
the MSD-TAMSD. In this section, we provide two examples: RL-FBM and L{\'e}vy
walks, both widely used stochastic models for biophysical systems. Despite
the nonergodicity of the process itself, the increments are asymptotically
stationary and thus ergodic at long times.

\subsection{Riemann-Liouville fractional integral representation of
fractional Brownian motion}

Recently, another variant of the fractional Langevin equation, namely, the fractional Langevin equation far from equilibrium (FLEFE)
\begin{eqnarray}
\frac{d^\alpha x(t)}{dt^\alpha} = \sqrt{2K_{\alpha}}\xi(t)
\end{eqnarray}
with $\alpha > 0$ was studied \cite{eab2011,wei2025}.
The initial conditions are specified as $x^{(k)}(0)=0$ for $k=0,\ldots,[\alpha]$, where $[\cdot]$ denotes the integer part of a positive real number.
$\xi(t)$ here represents
white Gaussian noise with zero mean and ACVF $\left<\xi(t)\xi(t')\right>=
\delta(t-t')$. The fractional derivative is to be interpreted in the Caputo
sense with order $n-1<\alpha<n$, where $n\in\mathbb{N}^+$ \cite{caputo1966,
caputo1967},
\begin{equation}
\label{caputo-derivative}
\frac{d^\alpha x(t)}{dt^\alpha}\equiv{}^C_0D^{\alpha}_t[x(t)]=\int_0^tx^{(n)}(u)
\frac{(t-u)^{-\alpha+n-1}}{\Gamma(n-\alpha)}du.
\end{equation} 
The conventional fractional Langevin equation originally suggested by Mori and
Kubo~\cite{mori1965,kubo1966,zwanzig2001,lutz2001} provides a stochastic
description of thermalized systems near equilibrium and thus obeys the
fluctuation-dissipation theorem (FDT) \cite{kubo1966}. In contrast, the FLEFE
does not obey the FDT, making it a more suitable candidate for evaluating the
observed stochastic dynamics in systems out of equilibrium, including the
movement of birds and other animals, or active particles such as bacteria
and amoeba; or financial market fluctuations and climate dynamics. 

The solution of the FLEFE is identical to FBM defined in terms
of a Riemann-Liouville fractional integral (RL-FBM) with an initial condition $x(0)=0$ and with Hurst exponent $H=
\alpha-1/2$, given by \cite{eab2011,lim2001}
\begin{equation}
\label{eq-rlfbm}
x(t)=\sqrt{2K_{\alpha}}\int_0^t\frac{(t-t')^{\alpha-1}}{\Gamma(\alpha)}\xi(t')dt'.
\end{equation} The RL-FBM was first introduced by L{\'e}vy \cite{levy1953,mandelbrot1968}.

Unlike FBM, RL-FBM has nonstationary increments. Recently, a
variant of RL-FBM has found application in disordered systems with
heterogeneous environments, e.g, when the diffusion coefficient and Hurst
exponent are functions of time or space or even randomly chosen from certain
distributions \cite{balcerek2022,slezak2023,wang2023,sam2025}. RL-FBM with
time-dependent exponents has been successfully validated in terms of the MSD
and the power spectral density through recent experimental measurements of
quantum dot motion within the cytoplasm of live mammalian cells, as observed by
SPT \cite{balcerek2023}.

The MSD of RL-FBM (\ref{eq-rlfbm}) reads
\begin{equation}
\label{eq-rlfbm-msd}
\langle\Omega^2(t)\rangle=\frac{2K_{\alpha}}{(2\alpha-1)\Gamma(\alpha)^2}t^{
2\alpha-1}, 
\end{equation}
and thus one can see that the MSD is physically meaningful for all $\alpha>1/2$.

The exact MSI and EA-TAMSD can be expressed in terms of Fox $H$-functions
\cite{wei2025},
\begin{eqnarray}
\nonumber
\left<\delta^2(\Delta;t)\right>&=&\frac{2K_{\alpha}}{(2\alpha-1)\Gamma(\alpha)^2}
[(t+\Delta)^{2\alpha-1}+t^{2\alpha-1}]\\
&&\hspace*{-1.8cm}
-\frac{4K_{\alpha}t^{\alpha}\Delta^{\alpha-1}}{\Gamma(\alpha)^2\Gamma(1-\alpha)}
H_{2,2}^{2,1}\left[\frac{\Delta}{t}\left|\begin{array}{l}(1,1),(1+\alpha,1)\\
(1-\alpha,1),(\alpha,1)\end{array}\right.\right],
\label{eq-rlfbm-msi}
\end{eqnarray} 
and 
\begin{eqnarray}
\nonumber
\left<\overline{\delta^2(\Delta;T)}\right>&=&\frac{K_\alpha}{\alpha(2\alpha-
1)\Gamma(\alpha)^2}\\
\nonumber
&&\times\left[\frac{T^{2\alpha}-\Delta^{2\alpha}}{T-\Delta}+(T-\Delta)^{2\alpha
-1}\right]\\
\nonumber
&&\hspace*{-2.0cm}
-\frac{4K_\alpha}{\Gamma(\alpha)^2\Gamma(1-\alpha)}\frac{(T-\Delta)^{\alpha}}{
\Delta^{1-\alpha}}\\
&&\hspace*{-2.0cm}
\times H_{3,3}^{1,3}\left[\frac{T-\Delta}{\Delta}\left|\begin{array}{l}
(-\alpha,1),(\alpha,1),(1-\alpha,1)\\(0,1),(-1-\alpha,1),(-\alpha,1)\end{array}
\right.\right].
\label{eq-rlfbm-EA-TAMSD}
\end{eqnarray} 

From the exact expression (\ref{eq-rlfbm-msi}) for the MSI, we see that the
increments are nonstationary; however, at long times, the increments of RL-FBM
become asymptotically stationary and moreover ergodic. This is because both the MSI and EA-TAMSD
share the same asymptotic form for $1/2<\alpha<3/2$ and for both $t,T\gg\Delta$
\cite{wei2025},
\begin{equation}
\label{eq-rlfbm-msi-tamsd}
\left<\overline{\delta^2(\Delta;T)}\right>\sim\left<\delta^2(\Delta;t)\right>
\sim\frac{2K_{\alpha}}{\Gamma(2\alpha)|\cos(\pi\alpha)|}\Delta^{2\alpha-1}.
\end{equation} We also check by simulations the scatter of
the TAMSDs characterized by the EB parameter with fixed lag time equal to the
time step $\Delta=\delta t$ in Fig.~\ref{eb-rlfbm}. The results are the same
as for FBM with Hurst exponent $H=\alpha-1/2$, i.e., for $1/2<\alpha
<1$, $\mathrm{EB}(\Delta=\delta t;T)\simeq T^{-1}$ and for $1<\alpha<3/2$,
$\mathrm{EB}(\Delta=\delta t;T)\simeq T^{4\alpha-6}$. Therefore, the increments
are ergodic. 

\begin{figure}
\includegraphics[width=0.9\linewidth]{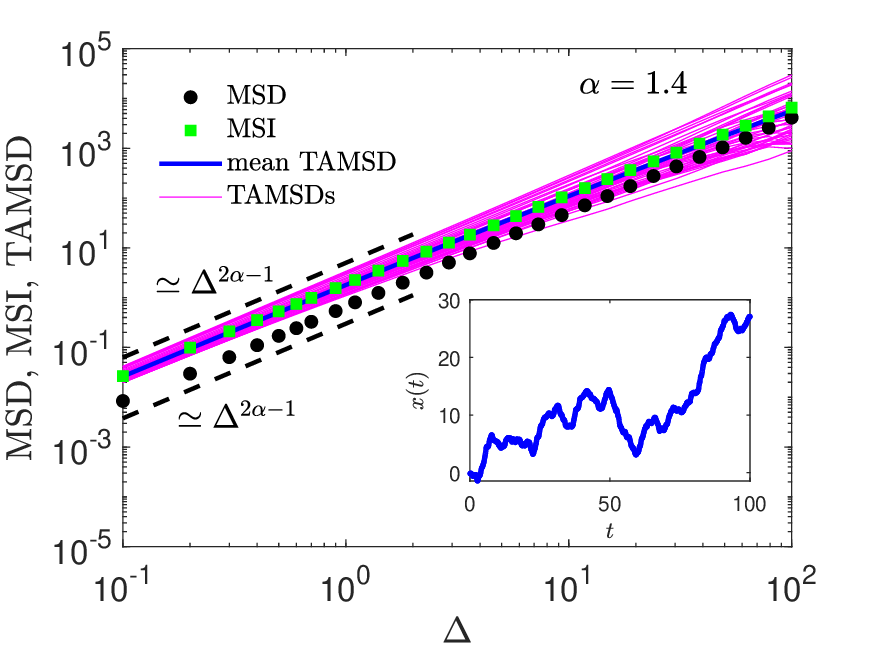}
\caption{Simulation results for RL-FBM (\ref{eq-rlfbm}) showing MSD (black circles), MSI
(green rectangles) and TAMSDs of individual trajectories (magenta curves) with
$\alpha=1.4$. The EA-TAMSD is represented by solid blue curves. When $1/2<
\alpha<3/2$, the EA-TAMSD of RL-FBM (\ref{eq-rlfbm}) is identical to the MSI
rather than the MSD, leading to a spurious nonergodicity. Other parameters:
time step $dt=0.01$, measurement time $T=500$, starting time of MSI $t=10$.
The algorithm of numerical simulations is presented in Appendix~C of
\cite{wei2025}.}
\label{fig-rlfbm}
\end{figure}

It is notable that in this regime $1/2<\alpha<3/2$, in the long time limit the
EA-TAMSD (\ref{eq-rlfbm-msi-tamsd}) differs from the MSD (\ref{eq-rlfbm-msd})
in the prefactor, as shown in Fig.~\ref{fig-rlfbm}. This effect was referred to
as ultraweak ergodicity breaking in Ref.~\cite{godec2013}. Instead, we see that
the EA-TAMSD converges to the MSI.

For $\alpha\ge3/2$, the increments of FLEFE motion (RL-FBM) are nonergodic at
all times $t>0$; however, higher order increments are asymptotically ergodic
with the $(n+1)$th order MSI ($n\ge1$) restoring stationarity in the regime
$(2n+1)/2<\alpha<(2n+3)/2$ \cite{wei2025}.

\subsection{L{\'e}vy walks}

The L{\'e}vy walk is a well-established stochastic model originally devised
in \cite{wong} (see also \cite{klablushle}), combining two key features, the
ability to generate anomalously faster motion than BM and a finite velocity of
the random walker, with broad applications in physics and biology \cite{gran2015,
fedo2018,klaf2015,geis1985,sagi2012,stan2014,hump2008,sims2014,zumofen1}.
In the simplest form the L{\'e}vy walks can be formulated as follows \cite{zumofen2} (see also \cite{kutner} for a more general case):
Consider a particle that alternates its velocity $v(t)$ randomly between the
values $+v_0$ and $-v_0$ with independent and identically distributed random
times $\tau$ ($0<\tau<\infty$) between these velocity changes. The switching
times are governed by the common PDF $\psi(\tau)$. To generate a sample of a
L{\'e}vy  walk, one can proceed as follows: The particle starts at $t=0$ with
velocity $+v_0$, travels a distance $v_0\tau_1$ where $\tau_1$ is drawn from
$\psi(\tau)$, before it is displaced by $-v_0\tau_2$, and so on. The position
of the particle becomes $x(t)=\int^t_0v(t')dt'$. L{\'e}vy walks in confined
domains and in external harmonic potentials were considered
\cite{bartek,pengbo}, including "soft" resetting in switching harmonic
confinement \cite{pengbo1}.

The L{\'e}vy walk is inherently a nonstationary process and therefore nonergodic
\cite{froemberg2013a,froemberg2013b,godec2013} and especially, the ultraweak
ergodicity breaking was observed in L{\'e}vy walks with power-law waiting time
PDF. Here, our focus is to examine the equivalence between the MSI and EA-TAMSD
in L{\'e}vy walks with exponential and power-law distributed waiting times. In
particular, this comparison may indicate that the increments of the process
become asymptotically stationary and ergodic even if the process itself remains
nonergodic.

\subsubsection{Exponentially distributed waiting times}

\begin{figure}
(a)\includegraphics[width=0.9\linewidth]{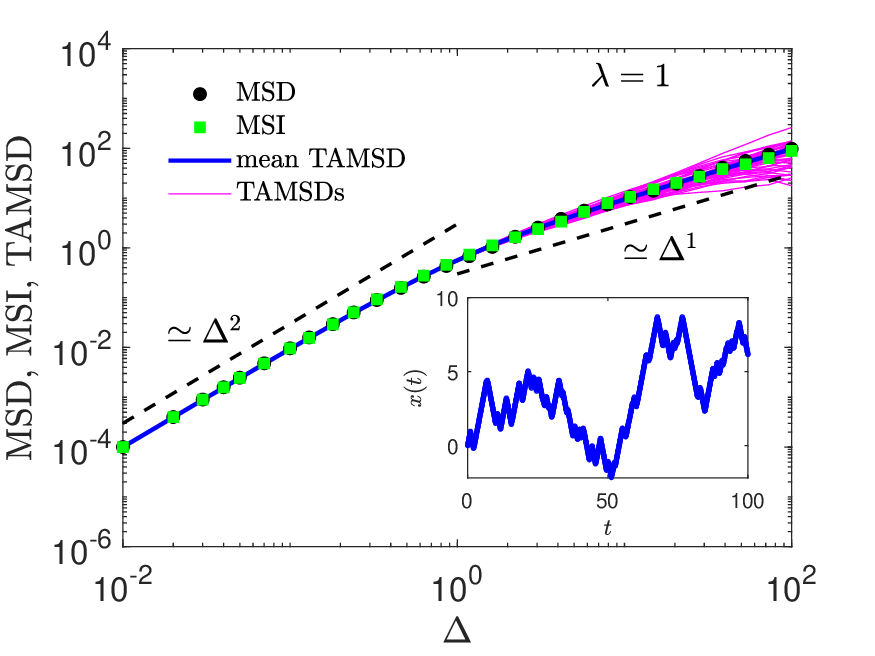}
(b)\includegraphics[width=0.9\linewidth]{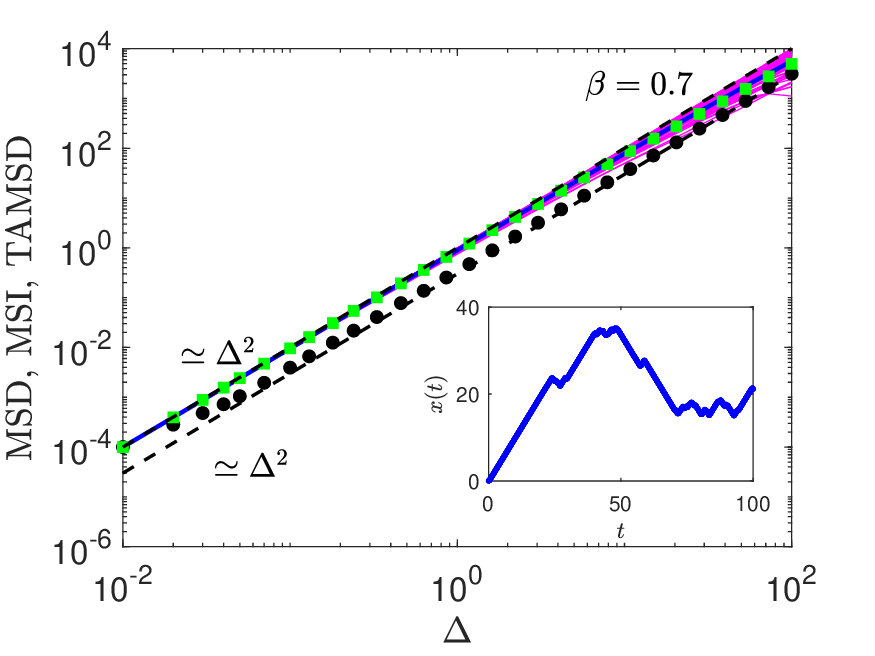}
(c)\includegraphics[width=0.9\linewidth]{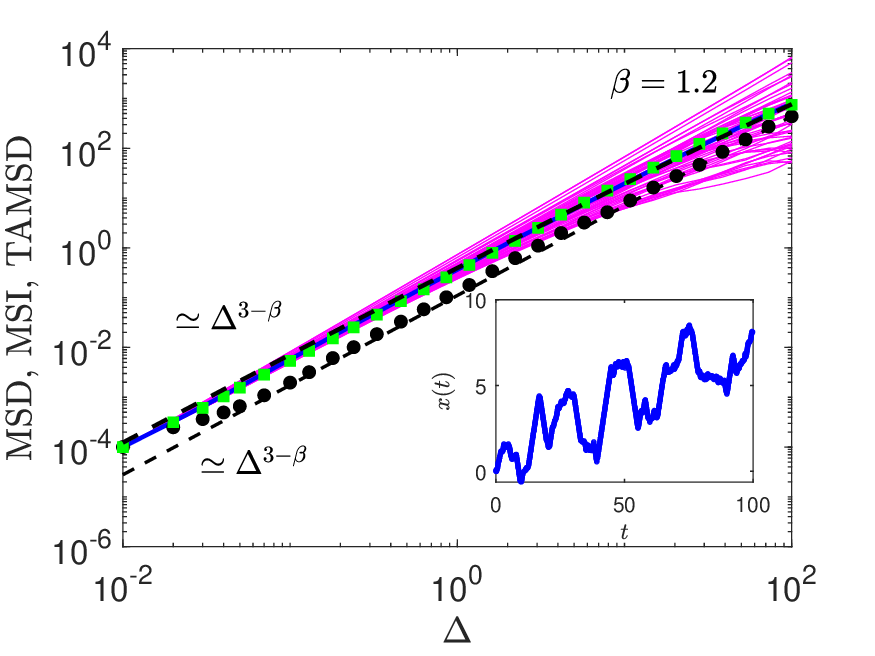}
\caption{Simulation results for L{\'e}vy walks showing the MSD (black circles),
MSI (green rectangles), and TAMSDs of 50 individual trajectories (magenta curves)
for (a) exponential waiting time PDF (\ref{expwtd}) with $\lambda=1$ and a
power-law waiting time PDF (\ref{plwtd}) with (b) $\beta=0.7$, and (c)
$\beta=1.2$. The EA-TAMSD is represented by solid blue curves. The variation
(scatter) of the amplitudes among individual TAMSDs
is negligible, indicating that individual trajectories are completely
reproducible. Other parameters: time step $d t=0.01$, measurement time $T=
500$, starting time of MSI $t=100$, and $\tau_0=0.01$. It is important to note that
discrepancies in the MSD arise as the power-law density cannot be accurately
generated in the short waiting time regime using the inverse transform sampling
method.}
\label{fig-lw}
\end{figure}

We first examine L{\'e}vy walks with exponentially distributed waiting times,
characterized by the waiting time PDF
\begin{equation}
\label{expwtd}
\psi(\tau)=\lambda\exp(-\lambda\tau),
\end{equation}
where $\lambda$ denotes the rate parameter. In this scenario, utilizing the
ACVF (\ref{eq-lw-acvf-exp}) of the position in Appendix~\ref{app-d}, the MSD,
MSI and EA-TAMSD all coincide,
\begin{eqnarray}
\label{eq-lw-exp-msd}
\nonumber
\left<\Omega^2(\Delta)\right>&=&\left<\delta^2(\Delta;t)\right>=\left<
\overline{\delta^2(\Delta;T)}\right>\\
&=&\frac{v_0^2}{2\lambda^2}[2\lambda\Delta+e^{-2\lambda\Delta}-1],
\end{eqnarray}
indicating that the increments of L{\'e}vy walks are stationary and,
furthermore, are ergodic. This is supported by Fig.~\ref{fig-lw}(a), which
shows consistent, reproducible TAMSDs. 

At short times $\Delta\ll\lambda^{-1}$, we obtain
\begin{equation}
\label{eq-lw-exp-st}
\left<\Omega^2(\Delta)\right>=\left<\delta^2(\Delta;t)\right>=\left<
\overline{\delta^2(\Delta;T)}\right>\sim v_0^2\Delta^2,
\end{equation}
thus revealing ballistic motion. In the long-time limit $\Delta\gg\lambda^{-1}$,
the second moments become
\begin{equation}
\label{eq-lw-exp-lt}
\left<\Omega^2(\Delta)\right>=\left<\delta^2(\Delta;t)\right>=\left<
\overline{\delta^2(\Delta;T)}\right>\sim\frac{v_0^2}{\lambda}\Delta,
\end{equation}
indicating normal-diffusive behavior.

Figure~\ref{fig-lw}(a) presents the identical MSI, MSD and EA-TAMSD for
L{\'e}vy walks with exponentially distributed waiting times. Notably, the dynamics exhibit a distinct crossover from ballistic diffusion at short times to normal diffusion at long times.

\subsubsection{Power-law distributed waiting times}

We now turn to the case when the random waiting times $\tau$ follow an
asymptotic power-law distribution as given by
\begin{equation}
\label{plwtd}
\psi(\tau)=\frac{\beta/\tau_0}{(1+\tau/\tau_0)^{1+\beta}},
\end{equation}
where $\tau_0$ is a constant. For $0<\beta<1$, the mean waiting time $\langle\tau
\rangle$ diverges, leading to ballistic motion of the process; for $1<\beta<2$,
the mean becomes finite, $\left<\tau\right>=\tau_0/(\beta-1)$, and the process is
superdiffusive \cite{klablushle,klaf2015}. 

In the ballistic motion regime $0<\beta<1$, the ACVF of the position can be
expressed in terms of the Beta function \cite{froemberg2013a,froemberg2013b},
as shown in Eq.~(\ref{eq-lw-x1}) in Appendix~\ref{app-d}. Consequently, the MSD
is given by 
\begin{equation}
\label{eq-lw-msd}
\left<\Omega^2(t)\right>=(1-\beta)v_0^2t^2,
\end{equation}
which characterizes ballistic motion.

In Appendix~\ref{app-d}, we utilize the connection between the Beta function and
the Fox $H-$function \cite{mathai2010}, along with the small-argument expansion
of the $H-$function, to derive the MSI for $t\gg\Delta$ as
\begin{eqnarray}
\nonumber
\left<\delta^2(\Delta;t)\right>&\sim&v_0^2\Delta^2\Bigg[1\\
&&\hspace*{-0.8cm}
+\frac{2\sin(\pi\beta)}{\pi(\beta-3)(\beta-2)(\beta-1)}\left(\frac{\Delta}{t}
\right)^{1-\beta}\Bigg],
\label{eq-lw-msi3}
\end{eqnarray}
and the EA-TAMSD for $T\gg \Delta$ as
\begin{eqnarray}
\nonumber
\left<\overline{\delta^2(\Delta;T)}\right>&\sim&v_0^2\Delta^2\Bigg[1\\
&&\hspace*{-1.2cm}
+\frac{2\sin(\pi\beta)}{\pi(\beta-3)(\beta-2)(\beta-1)\beta}\left(\frac{
\Delta}{T}\right)^{1-\beta}\Bigg].
\label{eq-lw-tamsd}
\end{eqnarray}
These results indicate that in the long-time limit, the increments of the
process become approximately stationary and depend predominantly on the lag
time. Moreover, the leading terms of the MSI and EA-TAMSD coincide and the
following relation holds
\begin{equation}
\label{eq-lw-pow1-msi-tamsd}
\left<\overline{\delta^2(\Delta;T)}\right>\sim\left<\delta^2(\Delta;t)\right>
\sim\frac{\left<\Omega^2(\Delta)\right>}{1-\beta}=v^2_0\Delta^2,
\end{equation}
where the leading term of the EA-TAMSD (\ref{eq-lw-tamsd}) differs from the MSD
(\ref{eq-lw-msd}) by the prefactor $1/(1-\beta)$, as also reported in
\cite{zumofen,godec2013,froemberg2013a}, demonstrating ultraweak ergodicity
breaking behavior according to the MSD-TAMSD criterion (\ref{eq-msd-tamsd}).
In contrast, we see that the leading term of the EA-TAMSD matches that of the
MSI~(\ref{eq-lw-msi3}), reinforcing the notion that the increments become asymptotically
stationary at long times. Additionally, the ergodicity of the increments is
further supported by the observation that the fluctuations of the TAMSDs vanish
in the long time limit, albeit slowly \cite{froemberg2013a,froemberg2013b}.
Interestingly, in the ballistic L{\'e}vy walk, nonergodicity in terms of
amplitude fluctuations of time averaged moments is not evident in the (decaying)
fluctuations of the TAMSD itself, but rather in the persistent fluctuations of
the shifted and rescaled dimensionless variable \cite{froemberg2013b}
\begin{equation}
\xi=\frac{v_0^2\Delta^2-\overline{\delta^2(\Delta;T)}}{v_0^2\Delta^2-\left<
\overline{\delta^2(\Delta;T)}\right>}.
\end{equation}

Figure~\ref{fig-lw}(b) presents simulation results illustrating the relationship
between MSD, MSI, and TAMSD in the ballistic L{\'e}vy walk case. It is important
to note that discrepancies observed in the simulated MSD stem from numerical inaccuracies in generating the power-law waiting time PDF (\ref{plwtd})
at short waiting time values using the inverse transform sampling method
\cite{kroese2011}.

For the superdiffusive motion regime $1<\beta<2$, the MSD can be obtained from
the ACVF provided in Eq.~(\ref{eq-lw-x2}) of Appendix~\ref{app-d}, resulting in
\begin{equation}
\label{eq-lw2-msd}
\left<\Omega^2(t)\right>=2K_{3-\beta}t^{3-\beta},  
\end{equation}
where $K_{3-\beta}$ is the generalized diffusion coefficient given by
\begin{equation}
K_{3-\beta}=\frac{\beta-1}{(3-\beta)(2-\beta)}v_0^2\tau_0^{\beta-1}.
\end{equation}
The exact expressions for the MSI and the EA-TAMSD are provided in
Eqs.~(\ref{eq-lw2-msi}) and (\ref{eq-lw2-tamsd}) in Appendix~\ref{app-d}. In
the limit $t\gg\Delta$, the MSI (\ref{eq-lw2-msi}) can be approximated as 
\begin{equation}
\label{eq-lw2-msi2}
\left<\delta^2(\Delta;t)\right>\sim\frac{2K_{3-\beta}}{\beta-1}\Delta^{3-\beta}
\Bigg[1-\frac{(3-\beta)(2-\beta)}{2}\left(\frac{\Delta}{t}\right)^{\beta-1}\Bigg],
\end{equation} indicating that the increments become asymptotically
stationary in the long-time limit,
while the EA-TAMSD~(\ref{eq-lw2-tamsd}) in the regime $T\gg\Delta$, takes on
the form
\begin{equation}
\label{eq-lw2-tamsd2}
\left<\overline{\delta^2(\Delta;T)}\right>\sim\frac{2K_{3-\beta}}{\beta-1}
\Delta^{3-\beta}\Bigg[1-\frac{3-\beta}{2}\left(\frac{\Delta}{T}\right)^{
\beta-1}\Bigg].
\end{equation}

Analogous to the ballistic regime, in the superdiffusive case and when $t,T\gg
\Delta$ we find that MSI and EA-TAMSD converge to the same expression, but
differ from the MSD (\ref{eq-lw2-msd}) by the prefactor $1/(\beta-1)$,
\begin{equation}
\label{eq-lw-pow2-msi-tamsd}
\left<\overline{\delta^2(\Delta;T)}\right>\sim\left<\delta^2(\Delta;t)\right>
\sim\frac{\left<\Omega^2(\Delta)\right>}{\beta-1}=\frac{2K_{3-\beta}}{\beta-1}\Delta
^{3-\beta}.
\end{equation}
This behavior is also confirmed by numerical simulations, as shown in
Fig.~\ref{fig-lw}(c). The EB parameter (\ref{eq-eb}) decays with
measurement time $T$, scaling as $T^{1-\beta}$
\cite{froemberg2013b}.

In summary, L{\'e}vy walks with power-law distributed waiting times, both
in the ballistic diffusion regime ($0<\beta<1$) and for superdiffusion ($1<
\beta<2$), exhibit asymptotically stationary increments at long
times, as evidenced by the asymptotic forms (\ref{eq-lw-msi3}) and
(\ref{eq-lw2-msi2}) of the MSI. Furthermore, the convergence of the EA-TAMSD to the MSI in the long-time limit points to the ergodic nature of the increments.
In contrast, comparisons between the MSD and TAMSD fail to capture this
fundamental property of the increments.

\section{Genuine Ergodicity breaking}
\label{section6}

The MSI-TAMSD criterion (\ref{eq-msi-tamsd}) not only reveals spurious
(non-)ergodicity in systems where the MSD-TAMSD criterion (\ref{eq-msd-tamsd})
fails, but is also effective for systems exhibiting genuine
ergodicity breaking. In this section, we
consider the CTRW, one of the most fundamental models describing anomalous
diffusion \cite{mont1965,metz2000,comp1996,hilfer1995,wong,klablushle,hughes}.

For the sake of simplicity, we consider a one-dimensional uncoupled CTRW, where random steps are separated by random waiting times. The jump lengths and waiting times are independent random variables, each drawn identically from their respective distributions, and the jump lengths are assumed to have finite variance. In the subordination framework \cite{feller,bochner}, the CTRW can be interpreted as the interplay of two distinct random processes, $x(t)=x(n(t))$. The parent process $x(n)$ represents a simple random walk with discrete steps, while the directing process (or subordinator) $n(t)$ acts as a renewal counting process that determines the random number of steps taken up to the physical time $t$. The associated
MSD and MSI satisfy \cite{soko2008,metz2014,soko2011}
\begin{equation}
\label{eq-ctrw-msd}
\left<\Omega^{2}(t)\right>=a^2\left< n(t)\right>,
\end{equation}
and 
\begin{equation}
\label{eq-ctrw-msi}
\left<\delta^2(\Delta;t)\right>=a^2\left[\left< n(t+\Delta)\right>-\left< n(t)
\right>\right],
\end{equation}
where $a^2$ is the mean-squared displacement in a single step.

\subsection{Exponentially-distributed waiting times}

When the waiting time follows the exponential PDF (\ref{expwtd}) with rate
$\lambda$, the mean of the counting process $n(t)$ takes the form
\cite{godreche2001,metz2014}
\begin{eqnarray}
\left<n(t)\right>=\lambda t.
\end{eqnarray}
Using this result, we obtain the equivalence between MSD, MSI, and EA-TAMSD,
\begin{equation}
\label{eq-ctrw-exp}
\left<\Omega^2(\Delta)\right>=\left<\delta^2(\Delta;t)\right>=\left<
\overline{\delta^2(\Delta;T)}\right>=2K_1\Delta,
\end{equation}
with the diffusion coefficient $K_1=\lambda a^2/2$. Figure~\ref{fig-ctrw}(a)
depicts simulation results for the MSD, MSI, and TAMSDs of the CTRW with
exponential waiting times. We note that the MSD, MSI and TAMSD indeed
coincide with each other, underlining its ergodic nature. It is important to
note that in
the continuum limit such a CTRW---with exponential waiting times and finite-variance jump
lengths---shares the same statistical properties as standard Brownian motion.

\subsection{Power-law distributed waiting time}

We now consider the case when the waiting time follows the power-law PDF
(\ref{plwtd}). For $0<\beta<1$, the mean waiting time $\langle\tau\rangle$
diverges, while for $1<\beta<2$, it becomes finite, $\langle \tau
\rangle=\tau_0/(\beta-1)$.

In the scale-free case $0<\beta<1$, the mean of the counting process $n(t)$
with $t\gg\tau_0$  is given by \cite{godreche2001,sokolov2001} 
\begin{equation}
\label{eq-ctrw-count}
\left<n(t)\right>\sim\frac{\sin(\pi\beta)}{\pi\beta}\left(\frac{t}{\tau_0}\right)
^\beta.
\end{equation}
Accordingly, the MSD scales as
\begin{equation}
\label{eq-ctrw-pow1-msd}
\left<\Omega^2(\Delta)\right>\sim 2K_\beta \Delta^\beta,
\end{equation}
where the generalized diffusion coefficient is given by $K_\beta=a^2\sin(\pi
\beta)/(2\pi\beta\tau_0^\beta)$.

In the long time limit $t\gg\Delta$ and $T\gg\Delta$ the MSI and EA-TAMSD
are, respectively,
\begin{eqnarray}
\nonumber
\left<\delta^2(\Delta;t)\right>&\sim&2K_\beta\left[\left(t+\Delta\right)^\beta
-t^\beta\right]\\
&\sim& 2\beta K_\beta\frac{\Delta}{t^{1-\beta}},
\label{eq-ctrw-pow1-msi}
\end{eqnarray}
and
\begin{eqnarray}
\nonumber
\left<\overline{\delta^2(\Delta;T)}\right>&\sim&\frac{2K_\beta}{\beta+1}\frac{
T^{\beta+1}-\Delta^{\beta+1}-(T-\Delta)^{\beta+1}}{T-\Delta}\\
&\sim&2K_\beta \frac{\Delta}{T^{1-\beta}}.
\label{eq-ctrw-pow1-tamsd}
\end{eqnarray}

\begin{figure}
\centering
(a)\includegraphics[width=1\linewidth]{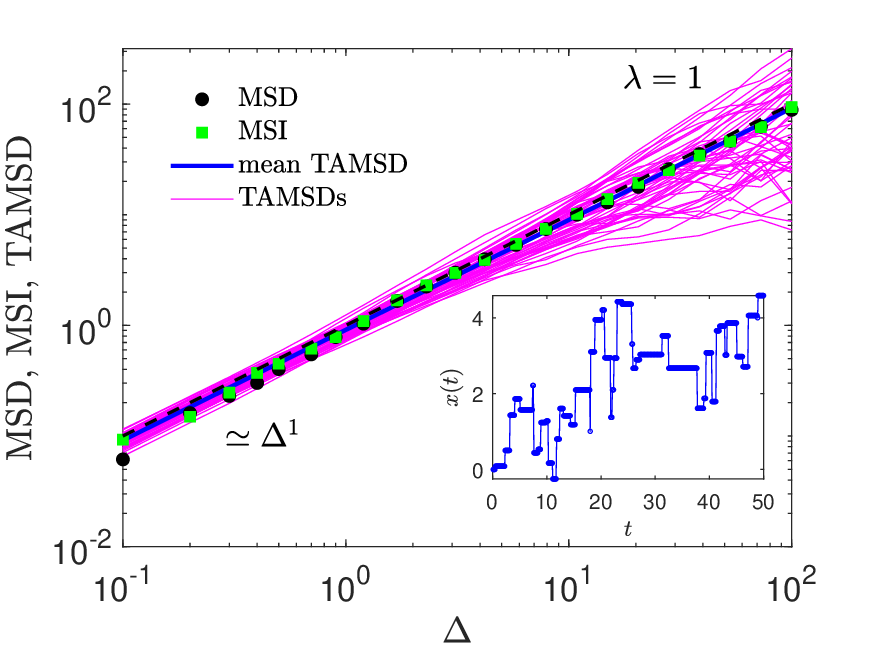}
(b)\includegraphics[width=1\linewidth]{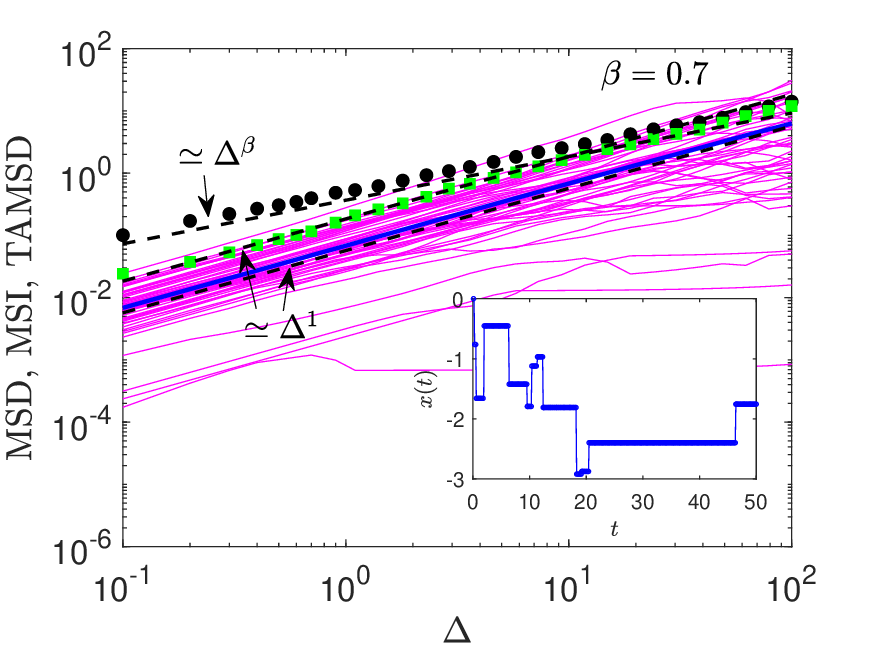}
(c)\includegraphics[width=1\linewidth]{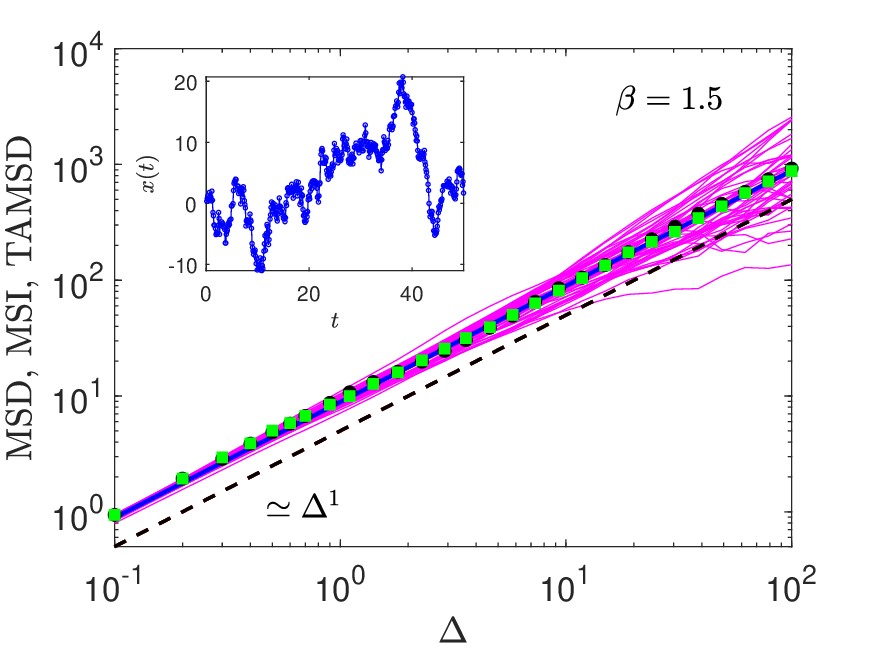}
\caption{Simulation results for CTRW showing MSD (black circles), MSI (green
rectangles), and TAMSDs from 50 individual trajectories (magenta curves), for
(a) exponential waiting time PDF (\ref{expwtd}) with $\lambda=1$ and power-law
waiting time PDF (\ref{plwtd}) with (b) $\beta=0.7$ and (c) $\beta=1.5$. The
EA-TAMSD is represented by solid blue curves. For the ergodicity breaking
in panel (b), the amplitude variation (scatter) among the TAMSDs of individual
trajectories becomes significantly pronounced, indicating a complete lack of
reproducibility between trajectories. Other parameters: time step $d t=
0.1$, measurement time $T=500$, starting time of MSI $t=10$,
dimensionless time scale $\tau_0=0.01$, MSI in a single
step $a^2=1$.
}
\label{fig-ctrw}
\end{figure}

Results from simulations for the CTRW with power-law waiting times are shown
in Fig.~\ref{fig-ctrw}(b), illustrating that the MSD, MSI, and EA-TAMSD are
all distinct. However, both MSI and EA-TAMSD exhibit asymptotically linear
scaling with the lag time, with prefactors that decay over time. This
behavior indicates weak ergodicity breaking in CTRWs with scale-free,
power-law distributed waiting times. Furthermore, the higher-order increments remain nonstationary in this regime; details can be found in
Appendix~\ref{app-e}.

For $\beta>1$, the mean of the counting process $n(t)$ for $t\gg\tau_0$ becomes
\cite{anna2020}
\begin{eqnarray}
\left<n(t)\right>\sim(\beta-1)\frac{t}{\tau_0}.
\end{eqnarray}
In this case, the MSD, MSI, and EA-TAMSD all coincide,
\begin{eqnarray}
\label{eq-ctrw-pow2-msd-msi}
\left<\Omega^2(\Delta)\right>\sim\left<\delta^2(\Delta;t)\right>\sim
\left\langle\overline{\delta^2(\Delta;T)}\right\rangle\sim 2K_1\Delta,
\end{eqnarray}
where the diffusion coefficient is $K_1=a^2(\beta-1)/(2\tau_0)$. These results
are also demonstrated by the simulations in Fig.~\ref{fig-ctrw}(c). Compared
with the CTRW with exponentially distributed waiting times, the equivalence
between MSD, MSI, and EA-TAMSD holds only at times longer than $\tau_0$.

Moreover, a CTRW with heavy-tailed waiting times subject to an external force is described by the fractional Fokker–Planck equation \cite{mebakla,metz2000} and exhibits weak ergodicity breaking \cite{barkai2009}.

\begin{table*}
\centering
\begin{ruledtabular}
\begin{tabular}{|c|c|c|c|c|c|}
\textbf{Process} & \textbf{MSD} & \textbf{MSI} & \textbf{EA-TAMSD} & \parbox[c][1.2cm][c]{2cm}{\textbf{Ergod. of} \\ \textbf{process}} & \parbox[c][1.2cm][c]{2cm}{\textbf{Ergod. of} \\ \textbf{increment}} \\
\hline
\parbox[c][0.8cm][c]{2.5cm}{\textbf{FBM}} & $2K_{2H}\Delta^{2H}$, Eq.~(\ref{eq-ergodicity-bm}) 
& $2K_{2H}\Delta^{2H}$, Eq.~(\ref{eq-ergodicity-bm}) 
& $2K_{2H}\Delta^{2H}$, Eq.~(\ref{eq-ergodicity-bm})  & \textbf{No} & \textbf{Yes}\\
\hline
\parbox[c][0.8cm][c]{2.5cm}{\textbf{OUP}}
& $\left<x_0^2\right>+\frac{\sigma^2}{2\lambda}$, Eq.~(\ref{eq-ou-msd2})
& $\frac{\sigma^2}{\lambda}$, Eq.~(\ref{eq-ergodicity-ou})
& $\frac{\sigma^2}{\lambda}$, Eq.~(\ref{eq-ergodicity-ou}) & \textbf{Yes} & \textbf{Yes}\\
\hline
\parbox[c][0.8cm][c]{3cm}{\textbf{fractional}\\\textbf{OUP}}
& \parbox[c]{3cm}{$\left<x_0^2\right>+\frac{\sigma_{2H}^2\Gamma({2H}+1)}{2\lambda^{{2H}}}$,\\ Eq.~(\ref{eq-foup-msd2})}
& \parbox[c][0.8cm][c]{3cm}{$\frac{\sigma_{2H}^2}{\lambda^{{2H}}}\Gamma({2H}+1)$,\\ Eq.~(\ref{eq-foup-msi-tamsd})}
& \parbox[c][0.8cm][c]{3cm}{$\frac{\sigma_{2H}^2}{\lambda^{{2H}}}\Gamma({2H}+1)$,\\ Eq.~(\ref{eq-foup-msi-tamsd})} & \textbf{Yes} & \textbf{Yes}\\
\hline
\parbox[c][1cm][c]{2.5cm}{\textbf{Reset FBM}}
& \parbox[c][0.8cm][c]{3cm}{$\frac{2K_{2H}}{r^{2H}}\Gamma({2H}+1)$,\\ Eq.~(\ref{eq-reset-msd-lt})}
& \parbox[c][0.8cm][c]{3cm}{$\frac{4K_{2H}}{r^{2H}}\Gamma({2H}+1)$,\\ Eq.~(\ref{eq-reset-msi-tamsd2})}
& \parbox[c][0.8cm][c]{3cm}{$\frac{4K_{2H}}{r^{2H}}\Gamma({2H}+1)$,\\ Eq.~(\ref{eq-reset-msi-tamsd2})} & \textbf{Yes} & \textbf{Yes}\\
\hline
\parbox[c][1cm][c]{2.5cm}{\textbf{RL-FBM}}
&\parbox[c][0.8cm][c]{3cm}{$\frac{2K_{\alpha}}{(2\alpha-1)\Gamma(\alpha)^2}\Delta^{2\alpha-1}$, \\Eq.~(\ref{eq-rlfbm-msd})}
& \parbox[c][0.8cm][c]{3cm}{$\frac{2K_{\alpha}}{\Gamma(2\alpha)
|\cos(\pi\alpha)|}\Delta^{2\alpha-1}$,\\ Eq.~(\ref{eq-rlfbm-msi-tamsd})}
& \parbox[c][0.8cm][c]{3cm}{$\frac{2K_{\alpha}}{\Gamma(2\alpha)
|\cos(\pi\alpha)|}\Delta^{2\alpha-1}$,\\ Eq.~(\ref{eq-rlfbm-msi-tamsd})} & \textbf{No} & \textbf{Yes}\\
\hline
\parbox[c]{2.5cm}{\textbf{L{\'e}vy walk} \\ \textbf{exponential w.t.}}
& \parbox[c][1.8cm][c]{3cm}{short time: $v_0^{2}\Delta^2$, Eq.~(\ref{eq-lw-exp-st}) \\ long time: $\frac{v_0^2}{\lambda}\Delta$, Eq.~(\ref{eq-lw-exp-lt})}
& \parbox[c][1.8cm][c]{3cm}{short time: $v_0^{2}\Delta^2$, Eq.~(\ref{eq-lw-exp-st}) \\ long time: $\frac{v_0^2}{\lambda}\Delta$, Eq.~(\ref{eq-lw-exp-lt})}
& \parbox[c][1.8cm][c]{3cm}{short time: $v_0^{2}\Delta^2$, Eq.~(\ref{eq-lw-exp-st}) \\ long time: $\frac{v_0^2}{\lambda}\Delta$, Eq.~(\ref{eq-lw-exp-lt})} & \textbf{No} & \textbf{Yes}\\
\hline
\parbox[c][1.2cm][c]{2.5cm}{\textbf{L{\'e}vy walk} \\ \textbf{power-law w.t. $0<\beta<1$}}
& $(1-\beta)v_{0}^{2}\Delta^{2}$, Eq.~(\ref{eq-lw-msd})
& $v_0^2\Delta^2$, Eq.~(\ref{eq-lw-pow1-msi-tamsd})
& $v_0^2\Delta^2$, Eq.~(\ref{eq-lw-pow1-msi-tamsd}) & \textbf{No} & \textbf{Yes}\\
\hline
\parbox[c][1.2cm][c]{2.5cm}{\textbf{L{\'e}vy walk} \\ \textbf{power-law w.t. $1<\beta<2$}}
& $2K_{3-\beta} \Delta^{3-\beta}$, Eq.~(\ref{eq-lw2-msd})
& $\frac{2K_{3-\beta}}{\beta-1}\Delta^{3-\beta}$, Eq.~(\ref{eq-lw-pow2-msi-tamsd})
& $\frac{2K_{3-\beta}}{\beta-1}\Delta^{3-\beta}$, Eq.~(\ref{eq-lw-pow2-msi-tamsd}) & \textbf{No} & \textbf{Yes}\\
\hline
\parbox[c][1.2cm][c]{2.5cm}{\textbf{CTRW} \\ \textbf{exponential w.t.}}
& $2K_1\Delta$, Eq.~(\ref{eq-ctrw-exp})
& $2K_1\Delta$, Eq.~(\ref{eq-ctrw-exp})
& $2K_1\Delta$, Eq.~(\ref{eq-ctrw-exp}) & \textbf{No} & \textbf{Yes}\\
\hline
\parbox[c][1.2cm][c]{2.5cm}{\textbf{CTRW} \\ \textbf{power-law w.t. $0<\beta<1$}}
& $2K_\beta\Delta^\beta$, Eq.~(\ref{eq-ctrw-pow1-msd})
& $2\beta K_\beta\frac{\Delta}{t^{1-\beta}}$, Eq.~(\ref{eq-ctrw-pow1-msi})
& $2K_\beta\frac{\Delta}{T^{1-\beta}}$, Eq.~(\ref{eq-ctrw-pow1-tamsd}) & \textbf{No} & \textbf{No}\\
\hline
\parbox[c][1.2cm][c]{2.5cm}{\textbf{CTRW} \\ \textbf{power-law w.t. $1<\beta<2$}}
& $2K_1\Delta$, Eq.~(\ref{eq-ctrw-pow2-msd-msi})
& $2K_1\Delta$, Eq.~(\ref{eq-ctrw-pow2-msd-msi})
& $2K_1\Delta$, Eq.~(\ref{eq-ctrw-pow2-msd-msi}) & \textbf{No} & \textbf{Yes}\\
\end{tabular}
\end{ruledtabular}
\caption{Comparison of the statistical properties of stochastic models and
the (non-)ergodic behavior of the processes themselves and their increments.}
\label{tab}
\end{table*}

\section{Overview of the results and outlook}
\label{section7}

Anomalous diffusion has been widely deduced from SPT experiments and large
scale computer simulations. While normal, Brownian diffusion is universal in
its convergence to a Gaussian displacement PDF and the linear time dependence
of the MSD, anomalous diffusion may have multiple origins and is described in
terms of a multitude of stochastic processes \cite{soko2015,metz2014a,soko2012}. Starting
from measured or simulated data, an important aspect of stochastic time series
analysis is the task of classification, determining which anomalous stochastic
model process is indeed encoded in the data, as well as the regression of the
associated model parameters. The classification can be achieved by statistical
observables such as the MSD and TAMSD \cite{barkai2012,metz2014a,soko2015}, the
p-variation method \cite{pvar}, higher order moments or the mean maximal
excursion method \cite{vincent}, or by the scaling exponents $J$, $L$, and $M$
introduced by Mandelbrot (see below) \cite{vilk2022,aghion2021}. Moreover,
dedicated Bayesian methods \cite{samu,samu1} and different machine learning
approaches \cite{jamali2021, jamali2025, andi,andi2,henrik,henrik1,janusz,gorka,yael,pinholt} have
been developed for this classification and regression tasks. Here we proposed
to include the MSI as a central observable and studied its performance for
different anomalous stochastic processes.

In SPT studies, the measured positional time series provides the relevant
information on the system, and the MSD remains the key physical observable to
characterize anomalous diffusion. 
%A key question in such analyses is whether
%the measured dynamics is ergodic. 
The discrepancy between MSD and TAMSD in the
limit of long trajectories has been widely and effectively employed to measure
nonergodicity (or weak ergodicity breaking) in numerous such systems. However,
as argued here, the typically employed MSD-TAMSD
criterion (\ref{eq-msd-tamsd}) may lead to incorrect conclusions when applied to some particular systems.

For real data, finite trajectory lengths and limited sampling constitute additional practical limitations. We illustrate these effects using simulated OUP trajectories in Appendix~\ref{app-f}. A finite measurement time primarily affects the TAMSD: when the lag time $\Delta$ becomes comparable to the trajectory length $T$, the number of available increments for time averaging decreases, leading to stronger trajectory-to-trajectory fluctuations and possible deviations of the mean TAMSD from the MSI, as shown in Fig.~\ref{f1}. Limited sampling, on the other hand, mainly affects ensemble-averaged quantities, such as the MSD and MSI, producing visible statistical fluctuations when only a small number of trajectories is available, as shown in Fig.~\ref{f2}. By contrast, the mean TAMSD is usually less sensitive to limited sampling, because each single-trajectory TAMSD already involves a time average over many increments.

In addition,  heterogeneity may further complicate the interpretation of real data. For instance, different cells or nanoparticles may have different intrinsic properties, such as size, shape, surface characteristics, local environments, or mobilities. Such heterogeneity can generate an additional scatter of individual TAMSDs and make signatures of nonergodicity harder to identify. From the perspective of superstatistics \cite{beck2003}, the measured ensemble may be viewed as a mixture of subensembles with different effective diffusivities or dynamical parameters. Even if each subensemble is ergodic, the combined population can display broad distributions of single-trajectory observables and apparent discrepancies between ensemble- and time-averaged quantities. Thus, an observed mismatch between ensemble- and time-averaged observables in real data may originate not only from the stochastic dynamics itself, but also from heterogeneity and experimental variability.

The three major results of this work are as follows. First, we demonstrated
that the MSD-TAMSD criterion (\ref{eq-msd-tamsd}) may introduce a spurious
ergodicity when applied to nonstationary and nonergodic processes but with
stationary increments, e.g., BM and FBM, and in turn produce spurious weak
ergodicity breaking in stationary and ergodic systems, e.g., the OUP and
FBM under stochastic resetting. Second, we propose that the MSI is a more
suitable observable than the MSD to assess the (non)ergodic properties of a
stochastic process via comparison with the TAMSD. The new MSI-TAMSD criterion
(\ref{eq-msi-tamsd}) in fact coincides with the classic definition
(\ref{eq-ergodicity}) of ergodicity and reflects the true (non)ergodic
nature of the above systems where the MSD fails. Third, we found that the
MSI can recognize long-time stationarity and ergodicity of increments of the
process in systems exhibiting ultraweak ergodicity breaking, in which the MSD
and TAMSD have the same scaling but a different prefactor. The behavior of MSD, MSI and EA-TAMSD for the  $11$ specific models considered in the text and discussed above are summarized  in Tab.~\ref{tab}. These results are ready to be used in the statistical analysis of particular experimental situations. Our work has broad relevance beyond time-series analysis. The examples considered in the manuscript are therefore not intended as a collection of model-specific applications, but rather serve to establish the domains of validity of the MSD--TAMSD comparison and the circumstances under which the MSI--TAMSD comparison is required.

The distinction between the MSD--TAMSD and MSI--TAMSD
comparisons can be summarized as follows. For processes
with a stationary state, a mismatch between MSD and TAMSD
may simply reflect a nonstationary preparation of the
initial ensemble: the MSD retains information about the
initial condition, whereas the TAMSD probes increments
after relaxation. This is the mechanism behind the
spurious nonergodicity found for the OUPs. For nonstationary processes, such as FBM, lacking a stationary
distribution, the situation is different. The issue
is no longer whether the initial distribution coincides with a
stationary one, but whether the lag-time increment
process is stationary, or asymptotically stationary. Since the TAMSD is a time average of squared
increments, its natural ensemble counterpart is the MSI, not
the MSD. The MSI--TAMSD comparison therefore provides a
direct test of the ergodicity of increments, while the
MSD--TAMSD comparison remains a useful diagnostic of aging, initial-condition dependence, and
nonstationarity.

Recent SPT studies revealed that intracellular transport of endo- and exogenous
tracers of various sizes is often not only anomalous, but also heterogeneous in
time and space. This implies that a single diffusion coefficient or scaling
exponent of standard anomalous-diffusion models is insufficient to describe the
underlying physical phenomena \cite{waig2023,wagh2023,han2020,bo2021,wang2009,
wang2012,diego2020}. Two main classes of models have been proposed to address
such complexity. The first is the Diffusing Diffusivity (DD) model
\cite{chub2014,chec2017,jain2016,tyagi,krapf2024,seno2020,seno2021,akimoto2023,akimoto2016,burov2021}, in which the diffusion coefficient is
treated as a stochastic process with temporal correlations. Over short
timescales, the diffusion coefficient remains relatively stable due to
strong correlations, reflecting local behavior. Over longer timescales,
this memory fades, and the particle motion is described by an effective
diffusivity when exploring the entire available space. Consequently, the
DD model typically yields stationary and ergodic increments. The second
class assumes that either the diffusion coefficient or the anomalous
diffusion exponent is a random variable drawn from a distribution---for
example, FBM with a random diffusivity \cite{gianni2016,magd2019} or random
exponent \cite{wosz2025,balc2022,sam2025}. In these models, the increments
are stationary but nonergodic, because each trajectory maintains a fixed,
strongly correlated diffusivity or exponent. As a result, individual TAMSDs
remain inherently irreproducible.

Beyond physical and biological systems, our findings may also find applications in finance, particularly in models such as the Black–Scholes \cite{black1973} and Heston \cite{heston} frameworks. In these models, the asset price $S(t)$ follows $dS(t)/S(t)=\sqrt{\nu(t)}dW(t)$, where $\nu(t)$ denotes the price volatility. In the Black–Scholes model, the volatility is constant, whereas in the Heston model it follows a Cox–Ingersoll–Ross (CIR) process \cite{cox1985}, which guarantees non-negativity, similar in spirit to the DD model. The Heston model reproduces well several important stylized facts of low-frequency price data, namely, leverage effect, time-varying volatility, and fat tails \cite{euch2019}.
Recent high-frequency data analyses suggest that volatility exhibits rough behavior, with the log-volatility effectively following a fractional Brownian motion (FBM) across practical timescales \cite{rose2018}. Because asset prices grow exponentially, the standard MSI is not directly applicable; instead, a logarithmic MSI can be defined as \begin{eqnarray}
R^2(\Delta,t)=\left<\left(\log S(t+\Delta)- \log S(t) \right)^2\right>.
\end{eqnarray}

For a nonstationary process, the origin of time can no longer be
chosen arbitrarily. This raises the question of aging, that is, the explicit
dependence of physical observables on the time span $t_a$ between the original
preparation of the system and the start of the recording of data. Aging is
a key property of glassy systems \cite{rott05} but was also rationalized
in polymeric semiconductors \cite{schub13}, potassium channel motion in the
membrane of human embryonic kidney cells \cite{diego2011}, living MIN6
insulisoma cells \cite{tabei}, the surface of hippocampal neurons \cite{fox},
groundwater systems \cite{kirchner}, quantum dots \cite{bouch2003}, drug
molecules in silica slabs \cite{amanda}, as well as financial time series
\cite{chers2017,ritschel}. Aging effects can be studied via the aged MSD
\cite{metz2014}, the latter being identical to the MSI and aged TAMSD in
temporally or spatially heterogeneous processes \cite{safd2015,metz23}.

In both RL-FBM and L{\'e}vy walk, which exhibit the phenomenon of ultraweak
ergodicity breaking, the asymptotic stationarity of the increments over
long times is demonstrated by the MSI. One might wonder if this asymptotic
stationarity of the increments is universal in ultraweak ergodicity
breaking phenomena. The answer lies in the origin of anomalous diffusion.
The EA-TAMSD and MSD can be expressed as \cite{vilk2022,aghion2021}
$\left<\overline{\delta^2(\Delta;T)}\right>\simeq T^{2L+2M-2}\Delta^{2J}$
and $\left<\Omega^2(\Delta)\right>\simeq \Delta^{2H}$, where the exponent
$J$ represents long-range correlations ("Joseph effect"), $L$  is related
to the fat-tailed probability density of increments ("Noah effect"), and $M$
reflects nonstationarity (the "Moses effect"). These exponents satisfy the
relation $H=J+L+M-1$ \cite{chen2017,vilk2022}. In ultraweak ergodicity breaking,
only a constant factor between MSD and TAMSD is observed, leading to $H=J$,
hence $L+M=1$, suggesting that processes with aging ($M<1/2$) and heavy-tailed
PDF of increments ($L>1/2$) could induce similar scaling of MSD and TAMSD.
RL-FBM and L{\'e}vy walks are specific cases with asymptotically stationary
increments ($M=1/2$) and non-heavy tails ($L=1/2$).

\section{Conclusions}
\label{section8}

SPT approaches have found extensive applications in many domains of science, particularly in  biological systems 
on subcellular, cellular and organism levels. Drawing reasonable conclusions from  SPT measurements implies application of 
different statistical approaches to their analysis and classification, in order to place the process at hand into one of the broader classes of stochastic 
processes (or to show that it belongs to a yet unknown class) or for parameter regression. One of these approaches is based on the comparison of the  MSD of a particle from its initial position at $t=0$  with a moving time average of the 
squared displacement over a fixed time interval $\Delta$ (TAMSD). The beginning of this $\Delta$-interval, at time $t$, slides
over the entire available part of the trajectory of the process, $0\leq  t \leq T-\Delta$. Moreover, in some cases an additional averaging 
over the ensemble of trajectories is applied. 

The equality of the two means above is often interpreted as a sign of the ergodicity 
of the process, while their difference is interpreted as a breaking thereof. While the difference or equality of these means can 
be used as a signature of the process, allowing to place the process at hand into the class of random processes with similar properties, 
the interpretation of the corresponding finding as a sign of ergodicity (breaking) is at best vague. The reason for this ambiguity is the
indiscriminate application of the definition of the ergodicity for a single-time observable ($\mathcal{O}(x(t))$ in Eq.~(\ref{eq-ergodicity}) to two-time
observables (increments $x(t+\Delta)-x(t)$).

Ergodicity of a random process leads to the independence of the property of interest on initial conditions at long times, and is an 
asymptotic property. However, the MSD is an ensemble mean of the very first increment of the process, between $t=0$ 
(which may be the preparation time of the system, when the initial condition applies) and $t= \Delta$. It may keep an explicit dependence 
on the initial condition, and be different from the time average even for a process which is ergodic according to a precise mathematical 
definition, as we have shown considering a particular example of the OUP. 
Moreover, a pre-requisite of ergodicity of a random process is its stationarity, so that applying the criterion to a nonstationary 
process cannot lead to any strict conclusions on its ergodicity or breaking thereof.

In the present article we discuss the MSD-TAMSD criterion and propose a new ramification based on the comparison of the ensemble 
mean of the increments of the process (MSI) with the TAMSD. Different from the MSD, the MSI is averaged over both
the ensemble of motions, and the initial time $t$ of the measurement interval $(t, t+\Delta)$. For stationary processes, and 
for nonstationary processes with stationary (or asymptotically stationary) increments, the equality of MSI and TAMSD signifies 
the ergodicity of the \textit{increments} of the process. 

The proposed statistical measure of the MSI offers a robust and consistent framework for the evaluation of genuine (non-)ergodicity of increments in both stationary and
nonstationary processes. 

To show this, we consider several examples covering a broad field of processes typically used as models 
for biological motions, see Table \ref{tab} in which  we summarize the results about the behavior of the MSD, MSI and TAMSD. 

Our detailed discussion of (non-)ergodicity in the context of stochastic time series clarifies some ambiguities in the current literature and will be of immediate relevance for the statistical analysis of SPT data from experiment and simulations.

\begin{acknowledgments}
We thank Andrey Cherstvy for valuable discussions. R.M. acknowledges financial
support from the German Science Foundation (DFG, Grant No. 318763901, CRC Data Assimilation and ME
1535/22-1) and NSF-BMFTR CRCNS (Grant 2112862/STAXS). A.V.C. acknowledges the
BMFTR project PLASMA-SPIN Energy (Grant 01DK24006). 
\end{acknowledgments}

\section*{DATA AVAILABILITY}
The code and data used to generate the simulation results and figures are available at \cite{code}.

\appendix

\makeatletter
\@addtoreset{figure}{section}
\@addtoreset{table}{section}
\makeatother
\renewcommand{\thefigure}{\thesection\arabic{figure}}
\renewcommand{\thetable}{\thesection\arabic{table}}

\section{Ergodicity breaking parameter}
\label{app-a}

\begin{figure}
\includegraphics[width=1\linewidth]{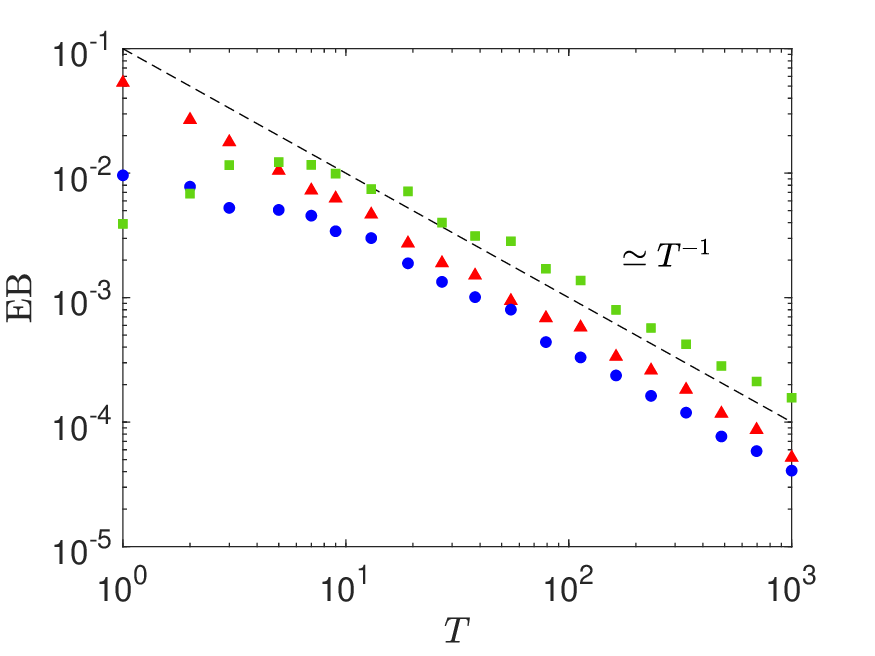}
\caption{Simulation results for the EB parameter of FBM under resetting for $H=0.2$ (red triangles), for $H=0.5$ (blue circles) and $H=0.8$ (green rectangles). The lag
time is fixed to the time step $\Delta=d t=0.01$, and the
resetting rate is $r=1$.}
\label{eb-reset}
\end{figure}

\begin{figure}
\includegraphics[width=1\linewidth]{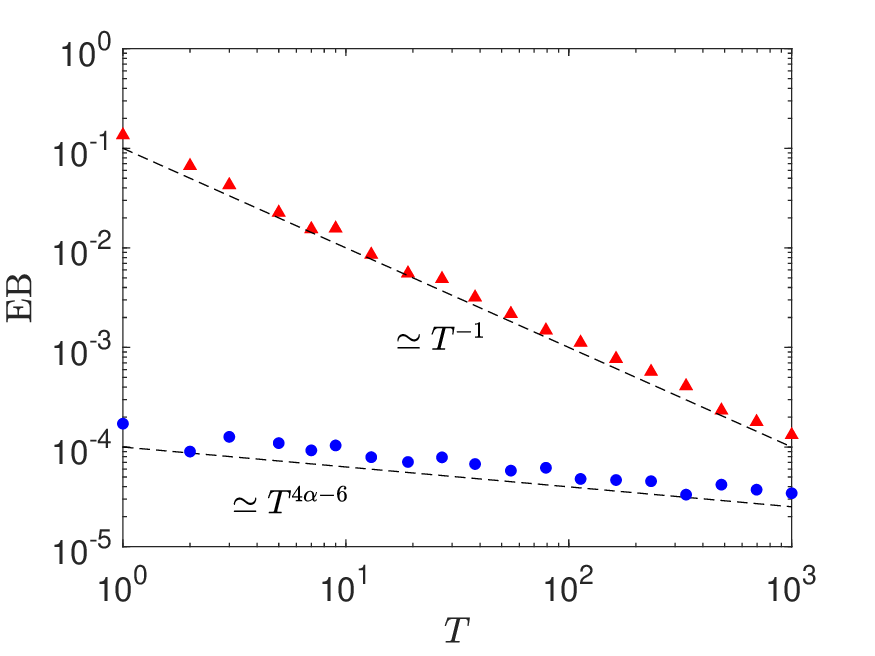}
\caption{Simulation results for the EB parameter $\mathrm{EB(\Delta;T)}$ of RL-FBM with $\alpha=1.4$
(blue circles) and $\alpha=0.7$ (red triangles).
The lag time is fixed to the time step  $\Delta=d t=0.01$.}
\label{eb-rlfbm}
\end{figure}

The randomness of each TAMSD of a stochastic process at a finite measurement
time $T$ gives rise to a certain amplitude scatter of the TAMSD around its
trajectory-to-trajectory average, the EA-TAMSD. This scatter can be quantified
by the EB parameter (\ref{eq-eb}) \cite{he2008,barkai2012,metz2014}. For
ordinary BM in the long time limit $T\to\infty$, one finds
\begin{equation}
\mathrm{EB}(\Delta;T)\sim\frac{4\Delta}{3T}.
\end{equation}
For FBM, the EB parameter for all $H$ is given by \cite{deng2009,maria} 
\begin{eqnarray}\label{eq-eb-fbm}
\mathrm{EB}(\Delta;T)\sim\left\{\begin{array}{ll}\displaystyle C_1\times\dfrac{
\Delta}{T},&0<H<3/4\\
C_2\times\left(\dfrac{\Delta}{T}\right)^{4-4H},&3/4<H<1\end{array}\right.,
\end{eqnarray}
where $C_1=\int_0^\infty[(1+s)^{2H}+|1-s|^{2H}-2s^{2H}]^2ds$ and $C_2=\left[2H(
2H-1)\right]^2\left[(4H-3)^{-1}-(4H-2)^{-1}\right]$. In the long-time limit $T\to\infty$, the EB parameter vanishes.
%In particular, the EB parameter is continuous at $H=3/4$ where a linear term with a logarithmic correction in $\Delta/T$ emerges \cite{godec2017}.
Analytical expressions for the EB parameters of RL-FBM and reset FBM are
currently unavailable. Therefore, we present simulation results instead. In
these simulations, we fix the lag time as the time step, $\Delta=d t$,
and vary the measurement time $T$. The analysis shows that for RL-FBM with
exponent $\alpha$, as shown in Fig.~\ref{eb-rlfbm}, the EB parameter exhibits the same long-time scaling as that of FBM (\ref{eq-eb-fbm}) with Hurst exponent $H=\alpha-1/2$.
For reset FBM in Fig.~\ref{eb-reset}, the EB parameter consistently shows an
asymptotic behavior proportional to $T^{-1}$ for all values of $H$.

\section{Fractional Ornstein-Uhlenbeck process}
\label{app-b}

The ACVF of the fractional OUP is given by \cite{jeon2012,mardoukhi2020}
\begin{widetext}
\begin{eqnarray} \label{eq-foup-acvf}
\nonumber
\left<x(t_1)x(t_2)\right>&=&\left<x_0^2\right>e^{-(t_1+t_2)}+\frac{\sigma_{
2H}^2}{2}\Bigg\{e^{-\lambda t_1}t_2^{2H}+e^{-\lambda t_2}t_1^{{2H}}-|t_1-t_2|
^{{2H}}\Bigg\}\\
\nonumber
&&+\frac{\sigma_{2H}^2}{4\lambda^{2H}}\Bigg\{e^{-\lambda|t_2-t_1|}\gamma(2H+1,
\lambda t_1)+e^{\lambda|t_2-t_1|}\gamma({2H}+1,\lambda t_2)-e^{\lambda|t_2-t_1|}
\gamma(2H+1,\lambda|t_2-t_1|)\Bigg\}\\
\nonumber
&&+\frac{\lambda\sigma_{2H}^2}{4({2H}+1)}\Bigg\{|t_2-t_1|^{2H+1}e^{-\lambda|
t_2-t_1|}\mathcal{M}({2H}+1,{2H}+2,\lambda|t_2-t_1|)\\
&&-t_1^{2H+1}e^{-\lambda (t_1+t_2)}\mathcal{M}(2H+1,2H+2,\lambda t_1)-t_2^{2H+1}
e^{-\lambda(t_1+t_2)}\mathcal{M}(2H+1,2H+2,\lambda t_2)\Bigg\},
\label{eq-foup-acvf}
\end{eqnarray}
\end{widetext}
where
\begin{eqnarray}
\gamma(z,x)=\int_0^xe^{-t}t^{z-1}dt   
\end{eqnarray}
is the lower incomplete gamma function and 
\begin{equation}
\mathcal{M}(a,b,z)=\frac{\Gamma(b)}{\Gamma(b-a)\Gamma(a)}\int_0^1 e^{z t}t^{a-1}
(1-t)^{b-a-1}dt.
\end{equation}
is the Kummer function (the confluent hypergeometric function of the first kind)
\cite{abramowitz1964}. Unlike the ACVF (\ref{eq-ou-acvf}) of the normal OUP which
becomes stationary when the initial condition is drawn from its equilibrium
distribution (\ref{eq-ou-distribution}), the ACVF of the fractional OUP remains
inherently nonstationary regardless of the initial condition chosen from the
stationary distribution (\ref{eq-foup-distribution}).

With the ACVF (\ref{eq-foup-acvf}), the MSD can be obtained as
\begin{eqnarray}
\label{eq-foup-msd-appendix}
\nonumber
\left<\Omega^2(t)\right>&=&\left<x^2_0\right>\left(1-e^{-\lambda t}\right)^2\\
\nonumber
&&+\frac{\sigma_{2H}^2}{2\lambda^{2H}}\gamma(2H+1,\lambda t)+\sigma_{2H}^2
t^{2H}e^{-\lambda t}\\
\nonumber
&&-\frac{\lambda\sigma_{2H}^2}{2(2H+1)}t^{2H+1}e^{-2\lambda t}\\
&&\times\mathcal{M}(2H+1,2H+2,\lambda t).
\end{eqnarray} 
For large $z$, the Kummer function can be expanded \cite{abramowitz1964},
\begin{eqnarray}
\nonumber
\mathcal{M}(a,b,z)&\sim&\frac{\Gamma(b)}{\Gamma(a)}e^zz^{a-b}\Big[1+\frac{(1-a)
(b-a)}{z}\\
&&\hspace*{-1.6cm}
+\frac{(1-a)(2-a)(b-a)(b-a+1)}{2z^2}+\dots\Big],
\end{eqnarray}
and for the incomplete gamma function we have
\begin{eqnarray}
\nonumber
\gamma(a,z)&\sim&\Gamma(a)-z^{a-1}e^{-z}\Big[1+\frac{a-1}{z}\\
&&+\frac{(a-1)(a-2)}{z^2}+\dots\Big].
\label{eq-gamma-expan}
\end{eqnarray}
We obtain the following approximation for the MSD (\ref{eq-foup-msd-appendix}),
\begin{equation}
\label{eq-foup-msd-app-st}
\left<\Omega^2(t)\right>\sim\left<x_{2H}^2\right>_{\mathrm{st}}-\frac{2\sigma^2
_{2H}}{\lambda^2}H(2H-1)t^{2H-2}e^{-\lambda t}, 
\end{equation}
which exponentially approaches the stationary
value  $\left<x_{2H}^2\right>_{\mathrm{st}}=\sigma_{2H}^2\Gamma({2H}+1)/(2
\lambda^{2H})$ with a power-law correction.

The exact expressions for the MSI and TAMSD for the fractional OUP are derived
in \cite{mardoukhi2020}. These expressions are quite complex, but they share
the same asymptotic form in the combined limit $t,T\gg\lambda^{-1}$, reading
\begin{eqnarray}
\label{eq-foup-msi-appendix}
\nonumber
\left<\delta^2(\Delta;t)\right>&\sim&\left<\overline{\delta^2(\Delta;T)}\right>\\
\nonumber
&&\hspace*{-2.0cm}
\sim\frac{\sigma_{2H}^2}{\lambda^{2H}}\Gamma(2H+1)+\sigma_{2H}^2 \Delta^{2H}\\
\nonumber
&&\hspace*{-1.6cm}
-\frac{\lambda\sigma_{2H}^2}{2(2H+1)}\Delta^{2H+1}e^{-\lambda\Delta}\mathcal{
M}(2H+1,2H+2,\lambda\Delta)\\
\nonumber
&&\hspace*{-1.6cm}
-\frac{\sigma_{2H}^2}{2\lambda^{2H}}\Big\{e^{\lambda\Delta}\Gamma(2H+1,\lambda
\Delta)+e^{-\lambda\Delta}\Gamma(2H+1)\Big\}.\\
\end{eqnarray}
where $\Gamma(a,z)=\int_z^\infty t^{a-1}e^{-t}dt$ is the upper incomplete gamma
function with the property $\Gamma(a,z)=\Gamma(z)-\gamma(a,z)$, where $\Gamma(z)$
is the complete gamma function \cite{abramowitz1964}. With the expansions of the
Kummer and the incomplete gamma function for $\Delta\gg\lambda^{-1}$, we obtain
\begin{eqnarray}
\nonumber
\left<\delta^2(\Delta;t)\right>&\sim&\left<\overline{\delta^2(\Delta;T)}\right>\\
&&\hspace*{-1.6cm}
\sim2\left<x_{2H}^2\right>_{\mathrm{st}}-\frac{2\sigma^2_{2H}}{\lambda^2}H(2H-1)
\Delta^{2H-2},
\label{eq-foup-msi-app-st}
\end{eqnarray}
approaching, in power-law form, the stationary value, when $H\neq1/2$.

\section{Fractional Brownian motion under resetting}
\label{app-c}

The ACVF of FBM with resetting is derived in \cite{majum2018}, from which the
MSI can be obtained. Alternatively, the MSI of the reset FBM can be directly
derived using a renewal approach \cite{chech2018, wang2022}, and it is given by
\begin{eqnarray}
\nonumber
\left<\delta^2(\Delta;t)\right>&=&2K_{2H}\bigg[e^{-r\Delta}\Delta^{2H}+\frac{
\gamma(2H+1,r\Delta)}{r^{2H}}\\
&&\hspace*{-1.8cm}
+\left(1-e^{-r\Delta}\right)\left(e^{-rt}t^{2H}+\frac{\gamma({2H}+1,rt)}{r^{2H}}
\right)\bigg].
\label{eq-reset-msi}
\end{eqnarray} 
The EA-TAMSD can then be obtained by integrating the MSI
\begin{eqnarray}
\nonumber
\left<\overline{\delta^2(\Delta;T)}\right>&=&2K_{2H}\bigg[e^{-r\Delta}\Delta^{
2H}+\frac{\gamma(2H+1,r\Delta)}{r^{2H}}\bigg]\\
\nonumber
&&+2K_{2H}\frac{1-e^{-r\Delta}}{r^{2H}}\bigg[\gamma(2H+1,r(T-\Delta))\\
\nonumber
&&+\frac{1}{r(T-\Delta)}\Big(\gamma(2H+1,r(T-\Delta))\\
&&-\gamma(2H+2,r(T-\Delta))\Big)\bigg].
\label{eq-reset-tamsd}
\end{eqnarray}
For $t$ and $T$ much larger than $r^{-1}$, both the MSI and the TAMSD exhibit
the same asymptotic behavior,
\begin{eqnarray}
\nonumber
\left<\delta^2(\Delta;t)\right>&\sim&\left<\overline{\delta^2(\Delta;T)}\right>\\
\nonumber
&\sim&2K_{2H}\Bigg[e^{-r\Delta}\Delta^{{2H}}+\frac{\gamma({2H}+1,r\Delta)}{
r^{2H}}\\
&&+\left(1-e^{-r\Delta}\right)\frac{\Gamma({2H}+1)}{r^{{2H}}}\Bigg],
\label{eq-reset-msi-tamsd}
\end{eqnarray}
At long lag times $\Delta\gg r^{-1}$, expanding the incomplete gamma function
according to Eq.~(\ref{eq-gamma-expan}), both the MSI and EA-TAMSD approach the
stationary value exponentially,
\begin{eqnarray}
\label{eq-reset-msi-app1}
\nonumber
\left<\delta^2(\Delta;t)\right>&\sim&\left<\overline{\delta^2(\Delta;T)}\right>\\
&&\hspace*{-1.2cm}
\sim4K_{2H}\Bigg[\frac{\Gamma(2H+1)}{r^{2H}}-\frac{H}{r}\Delta^{2H-1}e^{-r\Delta}\Bigg].
\end{eqnarray}
At short lag times $\Delta\ll r^{-1}$, with the approximation $\gamma(2H+1,r\Delta)
\sim(r\Delta)^{2H+1}/(2H+1)$ and $e^{-r\Delta}\sim1-r\Delta$, we have
\begin{eqnarray}
\label{eq-reset-msi-app2}
\left<\delta^2(\Delta;t)\right>&\sim&\left<\overline{\delta^2(\Delta;T)}\right>\nonumber\\
&\sim&2K_{2H}\left(\Delta^{2H}+\frac{\Gamma({2H}+1)}{r^{{2H}-1}}\Delta^1\right).
\end{eqnarray}
when $H<1/2$, the MSI and EA-TAMSD exhibit subdiffusive behavior at short times,
while for $H>1/2$ the linear term dominates, resulting in normal diffusion. In
the normal-diffusive case $H=1/2$, both terms contribute equally, leading to the
fact that MSI and TAMSD are approximately twice the MSD at short times.

\section{L{\'e}vy walks}
\label{app-d}

The displacement-ACVF of L{\'e}vy walks is related to the velocity-ACVF through
\begin{equation}
\label{eq-lw-acvf-d}
\left< x(t)x(t+\Delta)\right>=\int_0^tds_1\int_0^{t+\Delta}ds_2 \left<v(s_1)v(s_2)
\right>.
\end{equation} 
Since $v(t+\Delta)$ is either equal to $v(t)$ or $-v(t)$, the number $n$ of
directional changes within the time interval $[t,t+\Delta]$ must be even or
odd, accordingly. Therefore, the velocity-ACVF is given by
\begin{equation}
\label{eq-lw-acvf-v}
\left< v(t)v(t+\Delta)\right>=\sum_{n=0}^{\infty}(-1)^nv_0^2p_n(t,t+\Delta),
\end{equation}
where $p_n(t,t+\Delta)$ denotes the probability of observing $n$ velocity
reversals during the time interval $[t,t+\Delta]$.

\subsection{Exponentially-distributed waiting times}

For the case of the exponential PDF (\ref{expwtd}) of waiting times, the
probability of velocity reversals during $[t,t+\Delta]$ becomes
\cite{godreche2001}
\begin{equation}
\label{pn}
p_n(t,t+\Delta)=\frac{(\lambda\Delta)^n}{n!}e^{-\lambda\Delta}.
\end{equation}
Substituting Eq.~(\ref{pn}) into Eq.~(\ref{eq-lw-acvf-v}), we get
\begin{eqnarray}
\label{vv}
\nonumber
\left< v(t)v(t+\Delta)\right>&=&v_0^2\sum_{n=0}^{\infty}(-1)^np_n(t,t+\Delta)\\
\nonumber
&=&v_0^2\sum_{n=0}^{\infty}(-1)^n\frac{(\lambda\Delta)^n}{n!}e^{-\lambda\Delta}\\
&=&v_0^2e^{-2\lambda\Delta}.
\end{eqnarray}
Applying Eq.~(\ref{vv}) to Eq.~(\ref{eq-lw-acvf-d}), the displacement-ACVF is
given by
\begin{eqnarray}
\label{eq-lw-acvf-exp}
\nonumber
\left< x(t)x(t+\Delta)\right>&=&\frac{v_0^2}{(2\lambda)^2}\Big[4\lambda t+e^{-2
\lambda(t+\Delta)}\\
&&+e^{-2\lambda t}-e^{-2\lambda\Delta}-1\Big].
\end{eqnarray}
Then MSD, MSI, and EA-TAMSD can then be derived directly, as given by
Eq.~(\ref{eq-lw-exp-msd}) in the main text.

\vspace{8mm}

\subsection{Power-law distributed waiting times}

\subsubsection{Scale-free regime $0<\beta<1$}

For the ballistic motion regime with $0<\beta<1$, the velocity- and displacement-ACVFs are given by
\begin{equation}
\label{eq-lw-v1}
\left< v(t)v(t+\Delta)\right>=\frac{v_0^2\sin\pi\beta}{\pi}B\left(\frac{t}{t+
\Delta};\beta,1-\beta\right),
\end{equation}
and
\begin{widetext}
\begin{eqnarray}
\nonumber
\left< x(t)x(t+\Delta)\right>&=&\frac{v_0^2\sin\pi\beta}{2\pi}\Bigg[2t(t+\Delta)
B\left(\frac{t}{t+\Delta};\beta,1-\beta\right)-\left(t+\Delta\right)^2B\left(
\frac{t}{t+\Delta};1+\beta,1-\beta\right)\\
&&-t^2B\left(\frac{t}{t+\Delta};-1+\beta,1-\beta\right)\Bigg]-\frac{\beta(v_0
t)^2}{2},
\label{eq-lw-x1}
\end{eqnarray}
where $B(y;a,b)=\int_0^yu^{a-1}(1-u)^{b-1}du$ denotes the incomplete Beta
function \cite{abramowitz1964}.

The MSI can be exactly expressed as
\begin{eqnarray}
\nonumber
\left<\delta^2(\Delta;t)\right>&=&\frac{v_0^2\sin(\pi\beta)}{\pi}\Bigg[-2t(t+
\Delta)B\left(\frac{t}{t+\Delta};\beta,1-\beta\right)+(t+\Delta)^2B\left(\frac{
t}{t+\Delta};1+\beta,1-\beta\right)\\
&&+t^2 B\left(\frac{t}{t+\Delta};-1+\beta,1-\beta\right)\Bigg]+v_0^2t^2\left[1+
(1-\beta)\left(1+\frac{\Delta}{t}\right)^2\right].
\label{eq-lw-msi1}
\end{eqnarray}
In the first term of Eq.~(\ref{eq-lw-msi1}), using the relation $B(z;a,b)=\frac{
z^a}{a}{}_2F_1(a,1-b;a+1;z)$ between the incomplete Beta function and the
hypergeometric function $_2F_1$ along with the transformation between the
hypergeometric function and the Fox $H$-function, given by Eq.~(1.131) in
Ref.~\cite{mathai2010}, we find
\begin{eqnarray}
\nonumber
B\left(\frac{t}{t+\Delta};\beta,1-\beta\right)&=&\frac{1}{\beta}\left(\frac{t}{t
+\Delta}\right)^{\beta}\times{}_2F_1\left(\beta,\beta;\beta+1;\frac{t}{t+\Delta}
\right)\\
\nonumber
&=&\frac{1}{\beta}\left(\frac{t}{\Delta}\right)^{\beta}\times{}_2 F_1\left(\beta,
1;\beta+1;-\frac{t}{\Delta}\right)\\
&=&\left(\frac{t}{\Delta}\right)^{\beta}\times H_{2,2}^{1,2}\left[\frac{t}{
\Delta}\left|\begin{array}{l}(1-\beta,1),(0,1)\\(0,1),(-\beta,1)\end{array}
\right]\right.,
\end{eqnarray}
where in the second line, the Pfaff transformation ${}_2F_1(a,b;c;z)=(1-z)^{-
a}{}_2F_1(a,c-b;c;\frac{z}{z-1})$ \cite{abramowitz1964} was applied. Then,
using the identities (8.3.2.7) and (8.3.2.8) of the $H$-function in
\cite{prudnikov1990}, the incomplete Beta function can be expressed as a
special case of the $H$-function,
\begin{eqnarray}
\label{eq-beta1}
B\left(\frac{t}{t+\Delta};\beta,1-\beta\right)&=& H_{2,2}^{1,2}\left[\frac{t}{
\Delta}\left|\begin{array}{cc}(1,1),(\beta,1)\\(\beta,1),(0,1)\end{array}\right.
\right]=H_{2,2}^{2,1}\left[\frac{\Delta}{t}\left|\begin{array}{l}(1-\beta,1),(1,
1)\\(0,1),(1-\beta,1)\end{array}\right.\right].
\end{eqnarray}
Applying a similar transformation in the second term in the MSI
(\ref{eq-lw-msi1}), we obtain 
\begin{eqnarray}
\label{eq-beta2}
B\left(\frac{t}{t+\Delta}; 1+\beta,1-\beta\right)=H_{2,2}^{1,2}\left[\frac{t}{
\Delta}\left|\begin{array}{l}(1,1),(\beta,1)\\(1+\beta,1),(0,1)\end{array}
\right.\right]=H_{2,2}^{2,1}\left[\frac{\Delta}{t}\left|\begin{array}{l}
(-\beta,1),(1,1)\\(0,1),(1-\beta,1)\end{array}\right.\right].
\end{eqnarray}
The third term can be simplified via the identity $_2F_1(a,b;b;z)=(1-z)^{-a}$,
\begin{eqnarray}
\label{eq-beta3}
B\left(\frac{t}{t+\Delta};\beta-1,1-\beta\right)=\frac{1}{\beta-1}\left(\frac{t}{
t+\Delta}\right)^{\beta-1}{}_2F_1\left(\beta-1,\beta;\beta;\frac{t}{t+\Delta}
\right)=\frac{1}{\beta-1}\left(\frac{\Delta}{t}\right)^{1-\beta}.
\end{eqnarray}
Substituting Eqs.~(\ref{eq-beta1}) to (\ref{eq-beta3}) into the MSI
(\ref{eq-lw-msi1}), we find
\begin{eqnarray}
\label{eq-lw-msi2}
\nonumber
\left<\delta^2(\Delta;t)\right>&=&v_0^2\frac{\sin(\pi\beta)}{\pi}t^{2}\Bigg\{-
2\left(1+\frac{\Delta}{t}\right)H_{2,2}^{2,1}\left[\frac{\Delta}{t}\left|
\begin{array}{l}(1-\beta,1),(1,1)\\(0,1),(1-\beta,1)\end{array}\right.\right]
+\left(1+\frac{\Delta}{t}\right)^2H_{2,2}^{2,1}\left[\frac{\Delta}{t}\left|
\begin{array}{l}(-\beta,1),(1,1)\\(0,1),(1-\beta,1)\end{array}\right.\right]\\
&&+\frac{1}{\beta-1}\left(\frac{\Delta}{t}\right)^{1-\beta}\Bigg\}+v_0^2t^2
\left[1+(1-\beta)\left(1+\frac{\Delta}{t}\right)^2\right].
\end{eqnarray}

Using the expansion (8.3.2.3) of $H$-function in Ref.~\cite{prudnikov1990}
with $\Delta\ll t$, we have
\begin{eqnarray}
\nonumber
H_{2,2}^{2,1}\left[\frac{\Delta}{t}\left|\begin{array}{l}(1-\beta,1),(1,1)\\
(0,1),(1-\beta,1)\end{array}\right.\right]&=&\sum_{k=0}^{+\infty}\frac{(-1)^k
\Gamma(1-k-\beta)\Gamma(\beta+k)}{\Gamma(1-k)k!}\left(\frac{\Delta}{t}\right)^k\\
\nonumber
&&+\sum_{k=0}^{+\infty}\frac{(-1)^k\Gamma(\beta-1-k)\Gamma(k+1)}{\Gamma(\beta-k)
k!}\left(\frac{\Delta}{t}\right)^{1-\beta+k}\\
\nonumber
&=&\frac{\pi}{\sin(\pi\beta)}+\frac{1}{\beta-1}\left(\frac{\Delta}{t}\right)^{1-
\beta}-\frac{1}{\beta-2}\left(\frac{\Delta}{t}\right)^{2-\beta}\\
&&+\frac{1}{\beta-3}\left(\frac{\Delta}{t}\right)^{3-\beta}+O\left(\left(\frac{
\Delta}{t}\right)^{3-\beta}\right),
\label{h1}
\end{eqnarray}
where only the single term $k=0$ survives in the first summation since the Gamma
function has simple poles at non-positive integers. Likewise,
\begin{eqnarray}
\nonumber
H_{2,2}^{2,1}\left[\frac{\Delta}{t}\left|\begin{array}{l}(-\beta,1),(1,1)\\
(0,1),(1-\beta,1)\end{array}\right.\right]&=&\sum_{k=0}^{+\infty}\frac{(-
1)^k\Gamma(1-k-\beta)\Gamma(1+\beta+k)}{\Gamma(1-k)k!}\left(\frac{\Delta}{
t}\right)^k\\
\nonumber
&&+\sum_{k=0}^{+\infty}\frac{(-1)^k\Gamma(\beta-1-k)\Gamma(k+2)}{\Gamma(\beta-k)
k!}\left(\frac{\Delta}{t}\right)^{1-\beta+k}\\
\nonumber
&=&\frac{\pi\beta}{\sin(\pi\beta)} +\frac{1}{\beta-1}\left(\frac{\Delta}{t}
\right)^{1-\beta}-\frac{2}{\beta-2}\left(\frac{\Delta}{t}\right)^{2-\beta}\\
&&+\frac{3}{\beta-3}\left(\frac{\Delta}{t}\right)^{3-\beta}+O\left(\left(\frac{
\Delta}{t}\right)^{3-\beta}\right).
\label{h2}
\end{eqnarray}
Substituting Eqs.~(\ref{h1}) and (\ref{h2}) into the MSI (\ref{eq-lw-msi2}),
one can then see that the main order terms proportional to $t^2(\Delta/t)^0$,
$t^2(\Delta/t)^1$, $t^2(\Delta/t)^{1-\beta}$, and $t^2(\Delta/t)^{2-\beta}$
cancel out. With the main remaining terms, we derive the MSI (\ref{eq-lw-msi3})
for the long time limit $\Delta\ll t$,
\begin{eqnarray}
\nonumber
\left<\delta^2(\Delta;t)\right>&\sim&v_0^2t^2\left[\left(\frac{\Delta}{t}\right)
^2+\frac{2\sin(\pi\beta)}{\pi(\beta-3)(\beta-2)(\beta-1)}\left(\frac{\Delta}{t}
\right)^{3-\beta}+O\left(\left(\frac{\Delta}{t}\right)^{3-\beta}\right)\right]\\
\nonumber
&\sim&v_0^2\Delta^2\left[1+\frac{2\sin(\pi\beta)}{\pi(\beta-3)(\beta-2)(\beta-1
)}\left(\frac{\Delta}{t}\right)^{1-\beta}+O\left(\left(\frac{\Delta}{t}\right)^
{1-\beta}\right)\right]\\
&\sim&v_0^2\Delta^2.   
\end{eqnarray}   

Based on the MSI (\ref{eq-lw-msi2}), the exact EA-TAMSD can be expressed via
the Fox $H$-function as
\begin{equation}
\label{lw-tamsd1}
\left<\overline{\delta^2(\Delta;T)}\right>=\frac{v_0^2}{3}(T-\Delta)^2+\frac{
(1-\beta)v_0^2}{3}\frac{T^3-\Delta^3}{T-\Delta}+\frac{v_{0}^2\sin(\pi\beta)}{
\pi}\frac{\Delta^2}{T-\Delta}\int_0^{T-\Delta}I(t,\Delta)dt,
\end{equation}
where
\begin{eqnarray}
\nonumber
I(t,\Delta)&=&-2\frac{t}{\Delta}\left(1+\frac{t}{\Delta}\right)H_{2,2}^{1,2}
\left[\frac{t}{\Delta}\left|\begin{array}{l}(1,1),(\beta,1)\\(\beta,1),(0,1)
\end{array}\right.\right]+\left(1+\frac{t}{\Delta}\right)^2 H_{2,2}^{1,2}\left[
\frac{t}{\Delta}\left|\begin{array}{l}(1,1),(\beta,1)\\(1+\beta,1),(0,1)
\end{array}\right.\right]\\
&&+\frac{1}{\beta-1}\left(\frac{t}{\Delta}\right)^{\beta+1}.
\label{eq-function-I}
\end{eqnarray}
Using the integral (1.16.4.1) of the $H$-function and the identities (8.3.2.7)
and (8.3.2.8) in Ref.~\cite{prudnikov1990}, we can integrate the first
$H$-function in Eq.~(\ref{eq-function-I}),
\begin{eqnarray}
\nonumber 
&&\hspace*{-0.8cm}
\frac{1}{T-\Delta}\int_0^{T-\Delta}\frac{t}{\Delta}\left(1+\frac{t}{\Delta}
\right)H_{2,2}^{1,2}\left[\frac{t}{\Delta}\left|\begin{array}{l}(1,1),(\beta,1)\\
(\beta,1),(0,1)\end{array}\right.\right]dt\\
\nonumber 
&&=\frac{T-\Delta}{\Delta}H_{3,3}^{1,3}\left[\frac{T-\Delta}{\Delta}\left|
\begin{array}{l}(-1,1),(1,1),(\beta,1)\\(\beta,1),(0,1),(-2,1)\end{array}\right.
\right]+\left(\frac{T-\Delta}{\Delta}\right)^2 H_{3,3}^{1,3}\left[\frac{T-\Delta}{
\Delta}\left|\begin{array}{l}(-2,1),(1,1),(\beta,1)\\(\beta,1),(0,1),(-3,1)
\end{array}\right.\right]\\
\nonumber 
&&=H_{3,3}^{1,3}\left[\frac{T-\Delta}{\Delta}\left|\begin{array}{l}(0,1),(2,1),
(1+\beta,1)\\(1+\beta,1),(1,1),(-1,1)\end{array}\right.\right]+H_{3,3}^{1,3}
\left[\frac{T-\Delta}{\Delta}\left|\begin{array}{l}(0,1),(3,1),(2+\beta,1)\\
(2+\beta,1),(2,1),(-1,1)\end{array}\right.\right]\\
&&=H_{3,3}^{3,1}\left[\frac{\Delta}{T-\Delta}\left|\begin{array}{l}(-\beta,1),
(0,1),(2,1)\\(1,1),(-1,1),(-\beta,1)\end{array}\right.\right]+H_{3,3}^{3,1}
\left[\frac{\Delta}{T-\Delta}\left|\begin{array}{l}(-1-\beta,1),(-1,1),(2, 1)\\
(1,1),(-2,1),(-1-\beta,1)\end{array}\right.\right].
\end{eqnarray}
Applying similar steps in the second $H$-function, and using Eq.~(8.3.2.6) to
reduce to lower orders, one has 
\begin{eqnarray}
\nonumber 
&&\hspace*{-0.8cm}
\frac{1}{T-\Delta}\int_0^{T-\Delta}\left(1+\frac{t}{\Delta}\right)^{2}H_{2,2}^{
1,2}\left[\frac{t}{\Delta} \left|\begin{array}{l}(1,1),(\beta,1)\\(1+\beta,1),
(0,1)\end{array}\right.\right]dt\\
\nonumber 
&&=H_{2,2}^{2,1}\left[\frac{\Delta}{T-\Delta}\left|\begin{array}{l}(-\beta,1),
(2,1)\\(0,1),(1-\beta,1)\end{array}\right.\right]+2H_{3,3}^{3,1}\left[\frac{
\Delta}{T-\Delta}\left|\begin{array}{l}(-1-\beta,1),(0,1),(2,1)\\(1,1),(-1,1),
(-\beta,1)\end{array}\right.\right]\\
&&+H_{3,3}^{3,1}\left[\frac{\Delta}{T-\Delta}\left|\begin{array}{l}(-2-\beta,1),
(-1,1),(2,1)\\(1,1),(-2,1),(-1-\beta,1)\end{array}\right.\right].
\end{eqnarray}
Substituting the above results into the third term of Eq.~(\ref{lw-tamsd1})
yields
\begin{eqnarray}
\nonumber
\text{the third term}&=&\frac{\sin(\pi\beta)}{\pi}v_0^2\Delta^2\Bigg\{-2H_{3,3}
^{3,1}\left[\frac{\Delta}{T-\Delta}\left|\begin{array}{l}(-\beta,1),(0,1),(2,1)\\
(1,1),(-1,1),(-\beta,1)\end{array}\right.\right]\\
\nonumber
&&-2H_{3,3}^{3,1}\left[\frac{\Delta}{T-\Delta}\left|\begin{array}{l}(-1-\beta,1),
(-1,1),(2,1)\\(1,1),(-2,1),(-\beta-1,1)\end{array}\right.\right]\\
\nonumber
&&+H_{2,2}^{2,1}\left[\frac{\Delta}{T-\Delta}\left|\begin{array}{l}(-\beta,1),
(2,1)\\(0,1),(1-\beta,1)\end{array}\right.\right]+2H_{3,3}^{3,1}\left[\frac{
\Delta}{T-\Delta}\left| \begin{array}{l}(-1-\beta,1),(0,1),(2,1)\\(1,1),(-1,1),
(-\beta,1)\end{array}\right.\right]\\
&&+H_{3,3}^{3,1}\left[\frac{\Delta}{T-\Delta}\left|\begin{array}{l}(-2-\beta,1),
(-1,1),(2,1)\\(1,1),(-2,1),(-\beta-1,1)\end{array}\right.\right]+\frac{1}{(\beta
-1)(\beta+2)}\left(\frac{\Delta}{T-\Delta}\right)^{-1-\beta}\Bigg\}.
\end{eqnarray}

Finally, applying the expansion (8.3.2.3) of the $H$-function in
Ref.~\cite{prudnikov1990} with $\Delta\ll t$, and combining the first two terms
of Eq.~(\ref{lw-tamsd1}), one can see that the main order terms proportional to
$\Delta^2(\Delta/t)^{-2}$, $\Delta^2(\Delta/t)^{-1-\beta}$, $\Delta^2(\Delta/t)^
{-1}$, and $\Delta^2(\Delta/t)^{-\beta}$ cancel out. With the main remaining
terms, we can derive the TAMSD (\ref{lw-tamsd1}) in the long time limit,
\begin{eqnarray}
\nonumber
\left<\overline{\delta^2(\Delta;T)}\right>&\sim&v_0^2\Delta^2\Bigg[1+\frac{2\sin(
\pi\beta)}{\pi(\beta-3)(\beta-2)(\beta-1)\beta}\left(\frac{\Delta}{T}\right)^{1-
\beta}+O\left(\left(\frac{\Delta}{T}\right)^{1-\beta}\right)\Bigg]\\
&\sim&v_0^2\Delta^2.
\end{eqnarray}
\end{widetext}

\subsubsection{Regime $1<\beta<2$}

For the superdiffusive motion regime $1<\beta<2$, the velocity-ACVF is
\cite{froemberg2013a,froemberg2013b}
\begin{equation}
\label{v2}
\left< v(t)v(t+\Delta)\right>=\frac{v_0^2t_0^\beta}{(\beta-1)\langle\tau\rangle}
\left[\Delta^{1-\beta}-(t+\Delta)^{1-\beta}\right].
\end{equation}
Integrating Eq.~(\ref{v2}) as in Eq.~(\ref{eq-lw-acvf-d}), we obtain the
displacement-ACVF
\begin{eqnarray}
\label{eq-lw-x2}
\nonumber
\left< x(t)x(t+\Delta)\right>&=&\frac{K_{3-\beta}}{(\beta-1)}t^{3-\beta}\Bigg[
\beta+\left(1+\frac{\Delta}{t}\right)^{3-\beta}\\
&&\hspace*{-1.8cm}
-(3-\beta)\left(1+\frac{\Delta}{t}\right)^{2-\beta}-\left(\frac{\Delta}{t}
\right)^{3-\beta}\Bigg],
\end{eqnarray}
where
\begin{equation}
K_{3-\beta}=\frac{\beta-1}{(3-\beta)(2-\beta)}v_0^2t_0^{\beta-1}.
\end{equation}
The MSI can be obtained in the form
\begin{eqnarray}
\nonumber
\left<\delta^2(\Delta;t)\right>&=&\frac{2K_{3-\beta}}{\beta-1}t^{3-\beta}\Bigg[
-1-(2-\beta)\left(1+\frac{\Delta}{t}\right)^{3-\beta}\\
&&\hspace*{-1.2cm}
+(3-\beta)\left(1+\frac{\Delta}{t}\right)^{2-\beta}+\left(\frac{\Delta}{t}
\right)^{3-\beta}\Bigg].   
\label{eq-lw2-msi}
\end{eqnarray}
Expanding the expressions for $t\gg\Delta$, the MSI becomes
\begin{eqnarray}
\nonumber
\left<\delta^2(\Delta;t)\right>&\sim&\frac{2K_{3-\beta}}{\beta-1}t^{3-\beta}
\Bigg[\frac{(3-\beta)(2-\beta)}{2}\left(\frac{\Delta}{t}\right)^2\\
\nonumber
&&+\left(\frac{\Delta}{t}\right)^{3-\beta}+O\left(\left(\frac{\Delta}{t}\right)
^3\right)\Bigg]\\
\nonumber
&&\hspace*{-1.6cm}
\sim\frac{2K_{3-\beta}}{\beta-1}\Delta^{3-\beta}\\
&&\hspace*{-1.2cm}
\times\Bigg[1-\frac{(3-\beta)(2-\beta)}{2}\left(\frac{\Delta}{t}\right)^{\beta
-1}\Bigg].
\end{eqnarray}
Similarly, the exact EA-TAMSD can be derived as 
\begin{eqnarray}
\nonumber
\left<\overline{\delta^2(\Delta;T)}\right>&=&\frac{2K_{\beta}}{\beta-1}T^{3-
\beta}\Bigg[-\frac{1}{4-\beta}\left(1-\frac{\Delta}{T}\right)^{3-\beta}\\
\nonumber
&+&\frac{1}{4-\beta}\frac{1-(\Delta/T)^{4-\beta}}{1-\Delta/T}\\
&-&\frac{\Delta}{T}\frac{1-(\Delta/T)^{3-\beta}}{1-\Delta/T}
+\left(\frac{\Delta}{T}\right)^{3-\beta}\Bigg].
\label{eq-lw2-tamsd}
\end{eqnarray}
For $T\gg\Delta$, this can be approximated as
\begin{equation}
\left<\overline{\delta^2(\Delta;T)}\right>\sim\frac{2K_{3-\beta}}{\beta-1}
\Delta^{3-\beta}\Bigg[1-\frac{3-\beta}{2}\left(\frac{\Delta}{T}\right)^{\beta
-1}\Bigg].
\end{equation}

\section{CTRW $(n+1)$th order increment}
\label{app-e}

We consider the MSD in Eq.~(\ref{eq-msd}) as the zeroth order increment. The MSI defined in Eq.~(\ref{eq-msi}) is then identified as the first order increment, namely, \begin{equation}
\Delta^{(1)}x(t;\tau_1)=x(t+\tau_1)-x(t).
\end{equation}
The second order increment is defined as the increment of the first-order increment, \begin{equation}
\Delta^{(2)}x(t;\tau_1,\tau_2)=\Delta^{(1)}x(t+\tau_2;\tau_1)-\Delta^{(1)}x(t;
\tau_1).
\end{equation} With these definitions, the second-order increment can be interpreted as the
"velocity of the velocity".
By extension, the $(n+1)$th order increment is defined as the difference of the $n$th order
increment during the lag time $\tau_{n+1}$ \cite{schulz20131},
\begin{eqnarray}
\nonumber
\Delta^{(n+1)}x(t;\tau_1,\ldots,\tau_{n+1})&=&\Delta^{(n)}x(t+\tau_{n+1};\tau_1,
\ldots,\tau_n)\\
&-&\Delta^{(n)}x(t;\tau_1,\ldots,\tau_n),
\end{eqnarray}
similar in spirit to the theory of the $n$th order increment by Yaglom
\cite{yaglom1955,yaglom1953}. Consequently, it is reasonable to assume that $\tau
_n\gg\tau_{n-1}\gg\ldots\gg\tau_2\gg\tau_1$. 

%According to this definition, the first and second order increments are given
%by
%\begin{equation}
%\Delta^{(1)}x(t;\tau_1)=x(t+\tau_1)-x(t)
%\end{equation}
%and
%\begin{equation}
%\Delta^{(2)}x(t;\tau_1,\tau_2)=\Delta^{(1)}x(t+\tau_2;\tau_1)-\Delta^{(1)}x(t;
%\tau_1),
%\end{equation}
%where $\Delta^{(1)}x(t;\tau_1)$ is the first-order increment of the displacement
%and $\Delta^{(2)}x(t;\tau_1,\tau_2)$ is an increment of the first order increment.

Using the position-ACVF derived in \cite{baule2005},
\begin{equation}
\left<x(t_1)x(t_2)\right>=\Theta(t_2-t_1)\left<n(t_1)\right>+\Theta(t_1-t_2)
\left<n(t_2)\right>,
\end{equation}
where $\Theta(z)$ is the Heaviside step function, we find that the ACVF of the
first-order increment vanishes,
\begin{eqnarray}
\left<\Delta^{(1)}x(t+\Delta;\tau_1)\times\Delta^{(1)}x(t;\tau_1)\right>=0.
\end{eqnarray}
Then we obtain that the second order MSI in the regime $0<\beta<1$ takes on the
form 
\begin{eqnarray}
\nonumber
&&\left<\left[\Delta^{(2)}x(t;\tau_1,\Delta)\right]^2\right>\\
\nonumber
&&=\left<\left[\Delta^{(1)}x(t+\Delta;\tau_1)-\Delta^{(1)}x(t;\tau_1)\right]^2
\right\rangle\\
\nonumber
&&=\left<\left[\Delta^{(1)}x(t+\Delta;\tau_1)\right]^2\right>+\left<\left[
\Delta^{(1)}x(t;\tau_1)\right]^2\right>\\
\nonumber
&&=\left< n(t+\tau_1+\Delta)\right>-\left< n(t+\Delta)\right>\\
&&\hspace*{0.4cm}
+\left< n(t+\tau_1)\right>-\left<n(t)\right>.
\end{eqnarray}
Considering the mean (\ref{eq-ctrw-count}) of the counting process in the long
time limit $t\gg \Delta\gg\tau_1$,
\begin{eqnarray}
\nonumber
\left<\left[\Delta^{(2)}x(t;\tau_1,\Delta)\right]^2\right>&=&2K_\beta\Big[(t+
\tau_1+\Delta)^\beta\\
\nonumber
&&-(t+\Delta)^\beta+(t+\tau_1)^\beta-t^\beta\Big]\\
&\sim&4\beta K_\beta\left(\tau_1t^{\beta-1}\right),
\end{eqnarray} 
which is still nonstationary.

Moreover, the $(n+1)$th MSI can be generalized as
\begin{eqnarray}
\nonumber
&&\hspace*{-0.8cm}
\left<\left[\Delta^{(n+1)}x(t;\tau_1,\ldots,\tau_n,\Delta)\right]^2\right>\\
\nonumber
&&=\left<\Bigg[\Delta^{(n)}x(t+\Delta;\tau_1,\ldots,\tau_n)\right.\\
\nonumber
&&\left.-\Delta^{(n)}x(t;\tau_1,\ldots,\tau_n)\Bigg]^2\right>\\
&&\sim2n\beta K_\beta\left(\tau_1t^{\beta-1}\right),
\end{eqnarray}
in the long time limit $t\gg\tau_n\gg\ldots\gg\tau_1$, and they are all
nonstationary and dependent on $t^{\beta-1}$.

\section{Finite trajectory length and limited sampling effects for the OUP}\label{app-f}
\begin{figure}[!t]
\includegraphics[width=\columnwidth]{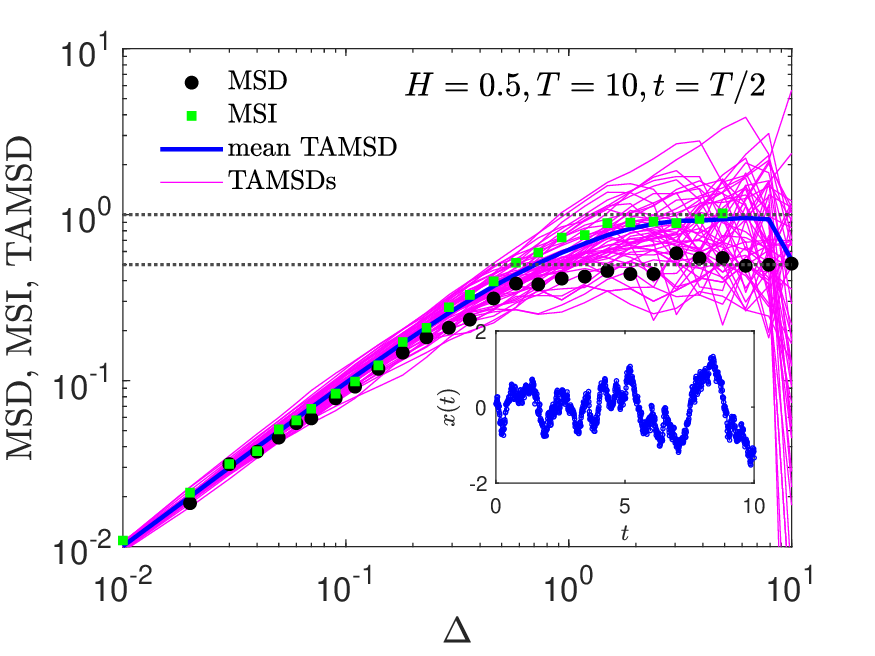}
\caption{Effects of finite trajectory length in the fractional OUP with
$H=0.5$,  measurement time $T=10$, and starting time for the MSI $t=T/2$. The MSD, MSI, ensemble-averaged TAMSD, and
individual TAMSDs are shown by black circles, green squares, the thick blue curve,
and thin magenta curves, respectively. The other parameters are the same as in Fig.~\ref{fig-foup}(a): initial condition $x_0=0$, time step $dt=0.01$, noise strength $\sigma_{2H}=1$, relaxation
rate $\lambda=1$ and number of trajectories $N=200$.}
\label{f1}
\end{figure}

First, we examine the effect of finite trajectory length by fixing the number of
trajectories and varying the measurement time $T$. 
For the TAMSD, the number of increments available for the time average decreases
as the lag time $\Delta$ increases. In the discrete case this number is
$(T-\Delta)/dt$.
Therefore, when $\Delta$ becomes comparable to the measurement time $T$,
the TAMSD is no longer a well-converged time average. This leads to enhanced
fluctuations of individual TAMSDs and may also bias
the ensemble-averaged TAMSD away from the MSI. In the limiting case $\Delta\to T$, the TAMSD is essentially determined by
only a few increments, and in the extreme case by the single displacement
$x(T)-x(0)$. Consequently, its ensemble average becomes closer to the MSD rather than to the MSI. This behavior is illustrated for the OUP with $H=0.5$ and measurement time
$T=10$ in Fig.~\ref{f1}. It explains the apparent downward bending of the mean
TAMSD and the large scatter of individual TAMSDs at long lag times in finite
trajectories. By contrast, the MSI is an ensemble-averaged increment observable and is not
affected by the reduction of the time-averaging window in the same way. Its
finite-time limitations are mainly that one must have $t+\Delta\leq T$, and
that the starting time $t$ should be sufficiently large for the process to
have reached the (asymptotically) stationary regime. Thus, finite measurement time
does not invalidate the MSI--TAMSD criterion, but it restricts its reliable
application to lag times satisfying $\Delta/T\ll1$.

\begin{figure}[!t]
\includegraphics[width=\columnwidth]{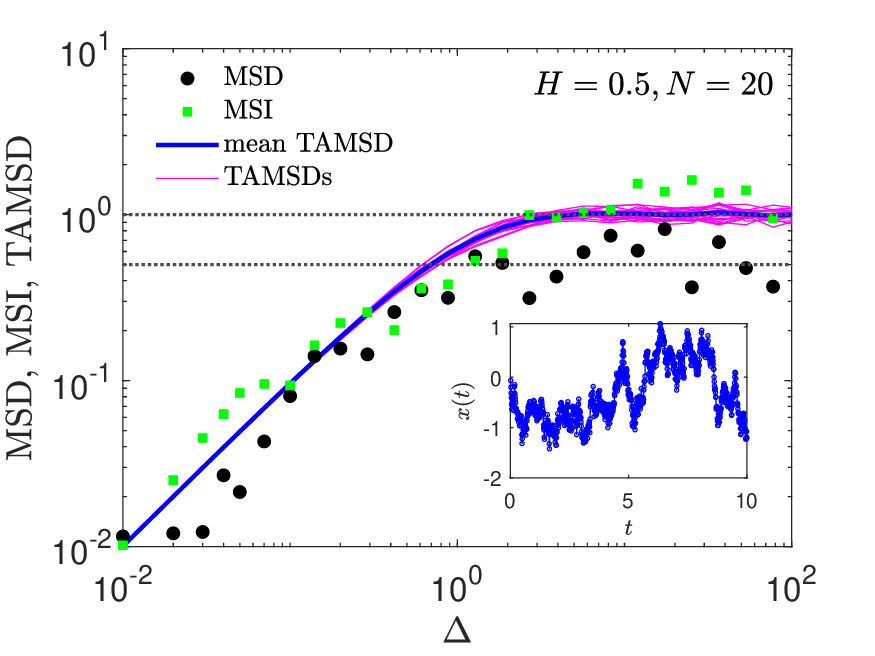}
\caption{Effects of limited sampling in the fractional OUP with
$H=0.5$ and number of trajectories $N=20$. The other parameters are the same as in Fig.~\ref{fig-foup}(a): initial condition $x_0=0$, time step $dt=0.01$, noise strength $\sigma_{2H}=1$, relaxation
rate $\lambda=1$, measurement time  $T=500$ and starting time for the MSI $t=10$.}
\label{f2}
\end{figure}

We also examined the influence of limited sampling within the same OUP model.
Limited sampling affects the MSD, MSI, and mean TAMSD in different ways. The
MSD and MSI are direct ensemble averages over squared displacements
or squared increments, and therefore they exhibit visible sampling fluctuations
when the number of trajectories is small. In contrast, the mean TAMSD is less
sensitive to limited sampling, because each individual TAMSD already involves
a time average over many increments along a single trajectory. As a result,
even for a relatively small ensemble, the mean TAMSD can remain rather smooth
and close to the MSI at small and intermediate lag times, as shown in
Fig.~\ref{f2}.

\end{document}